\documentclass{aa}

\usepackage{natbib}
\usepackage{graphicx}
\usepackage{txfonts}
\usepackage{amsmath}
\usepackage{orcidlink}
\usepackage{xcolor}
\definecolor{cobalt}{rgb}{0.06, 0.2, 0.65}
\usepackage{hyperref}
\hypersetup{colorlinks=true,
            citecolor=cobalt,
            linkcolor=cobalt,
            urlcolor=cobalt,
            pdftitle={Echoes from the dark: Galaxy catalog incompleteness in standard siren cosmology}}
            
\usepackage{multirow} 

\renewcommand{\d}[1]{\mathrm{d}#1}
\newcommand\given[1][]{\,#1\vert\,}
\newcommand{\vlambda}{\boldsymbol{\lambda}}
\newcommand{\vtheta}{\boldsymbol{\theta}}
\newcommand{\vdataGW}{\boldsymbol{d}}
\newcommand{\M}{M_\star}
\newcommand{\Mcut}{M_{\rm \star, cut}}

\defcitealias{Borghi2024}{B24}

\begin{document}

    \title{\vspace{-1em}Echoes from the dark:\\Galaxy catalog incompleteness in standard siren cosmology}
    \titlerunning{Echoes from the Dark}
    
    \author{Nicola~Borghi\inst{1,2,3}\orcidlink{0000-0002-2889-8997}\and
            Michele Moresco\inst{1,2} \orcidlink{0000-0002-7616-7136}\and
            Matteo Tagliazucchi\inst{1,2,3} \orcidlink{0009-0003-8886-3184}\and
            Giulia Cuomo\inst{4,5} \orcidlink{0009-0002-1094-3224}}
    
    \institute{Dipartimento di Fisica e Astronomia ``Augusto Righi''--Universit\`{a} di Bologna, via Gobetti 93/2, I-40129 Bologna, Italy\and
        INAF-OAS, Osservatorio di Astrofisica e Scienza dello Spazio di Bologna, via Gobetti 93/3, I-40129 Bologna, Italy\and
        INFN-Sezione di Bologna, Viale Berti Pichat 6/2, 40127 Bologna, Italy\and
        Dipartimento di Fisica ``G. Occhialini'', Universit\`a degli Studi di Milano-Bicocca, Piazza della Scienza 3, 20126 Milano, Italy\and
        INFN, Sezione di Milano-Bicocca, Piazza della Scienza 3, 20126 Milano, Italy\\
    }
    
    \abstract{
        Gravitational wave observations can be combined with galaxy catalogs to constrain cosmology and test modified gravity theories using the standard siren method. However, galaxy catalogs are intrinsically incomplete due to observational limitations, potentially leaving host galaxies undetected, thereby weakening constraints and potentially introducing systematic errors.
        In this work, we present a self-consistent framework to study catalog incompleteness and host weighting effects, implemented in the publicly available \texttt{CHIMERA} pipeline.
        We obtain joint cosmological and astrophysical population constraints from 100 binary black hole (BBH) events in a LIGO-Virgo-KAGRA O5-like configuration using spectroscopic galaxy catalogs with varying completeness levels and stellar-mass host weighting schemes.
        We find percent-level constraints on $H_0$ with complete catalogs, reaching precisions of $1.6\%$, $1.3\%$, and $0.9\%$ for constant, linear, and quadratic mass weighting, respectively.
        As completeness decreases, the precision degrades following a sigmoid trend, with a threshold and steepness that increase for stronger weightings. Simultaneously, the correlation between $H_0$ and the BBH population mass scale increases, making results more sensitive to assumptions about the astrophysical population.
        Remarkably, $2\%$ precision remains achievable even when catalogs contain only $50\%$ of the potential host galaxies within the gravitational wave detection horizon, while $1\%$ precision requires host probabilities scaling with stellar mass squared.
        The results are robust against host weighting mismodeling, even at moderate completeness levels.
        This work further highlights the importance of spectroscopic galaxy surveys in standard siren cosmology and provides a pathway for developing the science case of future facilities.
    }

\maketitle

\section{Introduction}\label{sec:intro}

Gravitational waves (GWs) from compact binary coalescences (CBCs) are maturing as a robust observational probe to constrain cosmology and fundamental physics, offering a new perspective on existing cosmological tensions \citep{Verde2019,Moresco2022,DiValentino2025}. Because GWs carry direct information on the CBC luminosity distance, they can be used as standard distance indicators or ``standard sirens'' \citep{Schutz1986,Holz2005}. With the additional source redshift information, it is then possible to constrain cosmology via the luminosity distance-redshift relation. However, the source redshift cannot be obtained directly from current GW data, as it is degenerate with the source masses.

When an electromagnetic counterpart is identified and the host galaxy redshift is measured, a ``bright siren'' measurement becomes possible. To date, the only confirmed case is GW170817 \citep{Abbott2017_GW170817}, constraining the Hubble constant to a precision of ${\sim}10\,\%$ \citep{Abbott2017_H0, Hotokezaka2019, Bulla2022, Palmese2024}. The latest release of the Gravitational-Wave Transient Catalog 4 (GWTC-4.0) by the LIGO-Virgo-KAGRA (LVK) Collaboration contains 219 confident CBC detections, the majority of which are binary black holes (BBHs) without electromagnetic counterparts \citep{GWTC1_Catalog,GWTC2_Catalog,GWTC3_Catalog,GWTC4_Catalog}. It is therefore essential to develop methods to extract the cosmological information without relying on counterpart detection.

These ``dark siren'' methods include two main complementary approaches. The first consists of statistically inferring the redshift from catalogs of potential host galaxies within the gravitational wave localization volume \citep{Schutz1986, DelPozzo2012, Chen2018, Fishbach2019_DarkGW170817, Gray2020, Palmese2020, Finke2021_DarkSirens, Gair2023}. The second consists of breaking the mass-redshift degeneracy by modeling intrinsic astrophysical properties. This is possible with ``spectral sirens,'' sources whose source-frame mass distribution contains features such as peaks, gaps, or changes in slope \citep{Chernoff1993,Taylor2012,Farr2019,Mastrogiovanni2021_icarogw,Mancarella2022,Ezquiaga2022,Mali2025}. Additional approaches rely on the cross-correlation between the spatial distribution of GWs and galaxies \citep[e.g.,][]{Oguri2016,Mukherjee2021,Ferri2025,Pedrotti2025}. 

Recent standard siren analyses \citep{Mastrogiovanni2023,Gray2023}, including the latest GWTC-4.0 cosmology analysis \citep{GWTC4_Cosmo}, have advanced beyond treating catalog and spectral siren methods as independent. Instead, they are jointly combined within a hierarchical Bayesian framework, making it possible to robustly marginalize over the uncertainties of the astrophysical and cosmological parameters simultaneously. Several pipelines are publicly available, including \texttt{gwcosmo} \citep{Gray2020, Gray2023}, \texttt{icarogw} \citep{Mastrogiovanni2023,Mastrogiovanni2024_icarogw}, and \texttt{CHIMERA} \citep{Borghi2024,Tagliazucchi2025}. For further details on the application of GWs to cosmology, we refer to \cite{Palmese2025} and \cite{Pierra2025}.

While galaxy catalogs provide valuable information, they are intrinsically limited by observational effects that can cause host galaxies to be missed, potentially weakening cosmological constraints \citep[e.g.,][]{Gray2020, Cross-Parkin2025} and introducing systematic errors \citep{Perna2025,Hanselman2025,Alfradique2025}. This ``incompleteness'' directly relates to assumptions about the CBC host galaxy population, as GW sources may preferentially reside in certain types of galaxies that are more easily detected and thus better represented in surveys. The link between CBCs and their host galaxies has been increasingly investigated through population synthesis studies \citep[e.g.,][]{Artale2020, Santoliquido2022} and observational analyses \citep{Vijaykumar2024}, although more BBH events are needed to establish robust observational constraints. As GW data quality improves, accounting for galaxy catalog incompleteness and weighting will become increasingly important. Extending current mock data analyses to model these effects and test for potential systematic errors in future standard siren measurements is a necessary step forward.

In our previous work \citep[][hereafter \citetalias{Borghi2024}]{Borghi2024}, we forecasted constraints on $H_0$ from 100 BBH mergers detected in an LVK O5-like scenario. We obtained percent-level constraints on $H_0$, while assuming a complete galaxy catalog with spectroscopic redshift measurements and considering only galaxies with stellar masses of $\log_{10}(\M/M_\odot) > 10.5$ as potential hosts. While this approximation was physically motivated by the expectation that BBHs form in more massive galaxies \citep[e.g.,][]{Artale2020,Santoliquido2022}, a more realistic approach requires including the full galaxy population in the analysis. Naively extending to all masses without additional weighting, however, would implicitly assume that low-mass $\log_{10}(\M/M_\odot){\approx}6$ galaxies have the same probability of hosting BBH mergers as massive $\log_{10}(\M/M_\odot) \approx 12$ ones. This strong assumption, which extends the BBH hosting probability over several orders of magnitude, requires accounting for BBH host weighting and observational effects in galaxy surveys.

In this paper, we present a self-consistent framework to study the effects of galaxy catalog incompleteness and host weighting in standard siren analyses, we incorporate it into a state-of-the-art analysis pipeline, and we provide updated and more realistic forecasts.
This framework can be readily applied to ongoing and upcoming surveys such as \emph{Euclid} \citep{EuclidCollaboration2025} and the Dark Energy Spectroscopic Instrument survey \citep[DESI;][]{DESICollaboration2024}, used to inform the science case of future facilities such as the Wide-Field Spectroscopic Telescope \citep[WST;][]{Mainieri2024}, and implemented in systematic studies for current cosmological analyses \citep{Agarwal2025}.
The paper is organized as follows. In Sect.~\ref{sec:method}, we introduce the statistical framework and its implementation in the \texttt{CHIMERA} code. In Sect.~\ref{sec:catalogs}, we describe the generation of mock galaxy and GW catalogs. Sect.~\ref{sec:results} provides the results of the incompleteness effects in standard siren constraints. Finally, in Sect.~\ref{sec:conclusions} we conclude the paper and discuss future applications of this method.

\section{Method}\label{sec:method}

We consider a population of GW sources, each described by the source-frame parameters, $\vtheta$ (such as redshift $z$, sky position $\hat{\Omega}$, and binary masses $m_1,\, m_2$), which globally follow a population distribution, $p_{\rm pop}(\vtheta \given \vlambda)$, described by hyperparameters, $\vlambda$. Given a dataset of $N$ independent GW detections, $\{\vdataGW_i\}$, the likelihood of $\vlambda$ can be obtained with the hierarchical Bayesian formalism \citep{Loredo2004,Mandel2019,Vitale2022}
\begin{equation}\label{eq:hyperlike}
    p(\{\vdataGW_i\}\given \vlambda) \propto \prod_{i=1}^{N} \frac{\int p(\vdataGW_i \given \vtheta, \vlambda)\, p_\mathrm{pop}(\vtheta \given \vlambda)\, \mathrm{d}\vtheta}{\xi(\vlambda)}\,,
\end{equation}
where $p(\vdataGW_i \given \vtheta, \vlambda)$ is the likelihood of individual GW events, and the selection function is defined as
\begin{equation}\label{eq:sel_func}
    \xi(\vlambda) = \int P_{\rm det}(\vtheta, \vlambda)\, p_{\rm pop}(\vtheta \given \vlambda)\, \mathrm{d}\vtheta,
\end{equation}
which corrects for the fraction of detectable events given a target population, with $P_{\rm det}(\vtheta, \vlambda)\in[0,1]$ being their detection probability. Equation~(\ref{eq:hyperlike}) assumes a scale-invariant rate prior so that the likelihood depends only on the shape of the population distribution, not on the overall rate normalization \citep{Mandel2019}.

It is crucial that $P_{\rm det}(\vtheta, \vlambda)$ is consistent with the detection pipeline. To ensure this, Eq.~(\ref{eq:sel_func}) is typically estimated by Monte Carlo integration, using a suite of simulated signals injected into the detection pipelines. This involves generating $N_\mathrm{gen}$ events with parameters $\theta_j$, then computing
\begin{equation}\label{eq:sel_func_est}
    \hat{\xi}(\vlambda) = \frac{1}{N_\mathrm{gen}} \sum_{i=1}^{N_\mathrm{det}} \frac{p_\mathrm{pop}(\vtheta_i|\vlambda)}{p_\mathrm{draw}(\vtheta_i)},
\end{equation}
where the summation runs over the $N_\mathrm{det}$ detected injections, reweighted by the ratio of the target population to the injection distribution.

\subsection{Population}
To model the BBH population, we separate the hyperparameters, $\vlambda$, into three categories: those that describe the underlying cosmology, $\vlambda_{\rm c}$; the mass distribution, $\vlambda_{\rm m}$; and the redshift distribution, $\vlambda_{\rm z}$. This allows us to write the population model as
\begin{equation}\label{eq:pop}
    p_{\rm pop}(\boldsymbol{\theta} \given \vlambda) = p(m_1, m_2 \given \vlambda_{\rm m})\; p(z, \hat{\Omega} \given \vlambda_{\rm c}, \vlambda_{\rm z}).
\end{equation}
The first term represents the intrinsic mass distribution, typically described using parametric models, although nonparametric approaches are also being studied \cite[e.g,][]{Edelman2022, Abbott2023_Population, Farah2025}. The second term acts as a ``redshift prior,'' which can be informed by external electromagnetic data, such as galaxy catalogs. Equation~(\ref{eq:pop}) assumes that the intrinsic BBH mass distribution does not evolve significantly on cosmological timescales, as supported by current data \citep[see][]{Callister2024_Review, GWTC4_Population}.
More generally, we can incorporate dependencies on
host galaxy properties, $\mathcal{G}$ (such as stellar mass, metallicity, and star formation rate), that may influence BBH formation, evolution, and ultimately their merger rate. In this case, the population prior is marginalized over galaxy properties,
\begin{equation}\label{eq:popG}
    p_{\rm pop}(\boldsymbol{\theta} \given \vlambda) = \int p_{\rm pop}(\boldsymbol{\theta}, \mathcal{G} \given \vlambda)\,\mathrm{d}\mathcal{G}\,.
\end{equation}
Here we do not consider observational uncertainties in galaxy properties to focus exclusively on incompleteness effects. We will address the inclusion of survey-dependent galaxy weight uncertainties in a subsequent work.

\subsection{Redshift prior}
The core of the methodology presented in this paper is the construction of a redshift prior informed by a galaxy catalog that includes host galaxy probability and completeness. The underlying assumption is that CBC events only occur inside galaxies. The correlation between BBHs and their host galaxy properties is an active area of research, and various theoretical models predict different dependencies based on formation channels, metallicity effects, and dynamical processes. In this work, we adopt a two-step approach to model these correlations: (i) we treat stellar mass as a property that varies on a galaxy-by-galaxy basis; and (ii) we model the BBH merger rate density as a parametric function in $z$. So we can rewrite the redshift prior of Eq.~(\ref{eq:popG}) as
\begin{equation}\label{eq:pz_galprop}
    p(z, \hat{\Omega}, \mathcal{G} \given \vlambda_{\rm c}, \vlambda_{\rm z}) \propto p_{\rm gal}(z, \hat{\Omega}, \mathcal{G} \given \vlambda_{\rm c})\,\frac{\psi(z \given \vlambda_{\rm z})}{(1+z)}\,,
\end{equation}
where $p_{\rm gal}(z, \hat{\Omega}, \mathcal{G} \given \vlambda_{\rm c})$ is the probability that a galaxy with properties $\mathcal{G}$ at redshift $z$ and sky location $\hat{\Omega}$ hosts a GW event, $\psi(z \given \vlambda_{\rm z})$ is the BBH merger rate density, and the factor $(1+z)^{-1}$ converts the source-frame time to the detector-frame time.

As a first step, we introduce the case for a complete galaxy catalog that contains all potential host galaxies. In this case, we can define $p_{\rm gal}(z, \hat{\Omega}, \mathcal{G} \given \vlambda_{\rm c}) = p_{\rm cat}(z, \hat{\Omega}, \mathcal{G} \given \vlambda_{\rm c})$
and compute this probability as a weighted mean over the contribution of each galaxy, $g$, in the catalog
\begin{equation}\label{eq:pcat}
    p_{\rm cat}(z, \hat{\Omega}, \mathcal{G} \given \vlambda_{\rm c})
    = \frac{\sum_g w_g\, p(z, \mathcal{G} \given \tilde{z}_g, \vlambda_\mathrm{c})\, \delta(\hat{\Omega}-\hat{\Omega}_g )}{\sum_g w_g}\,,
\end{equation}
where $w_g$ is the weight of the $g$-th galaxy, $p(z, \mathcal{G} \given \tilde{z}_g, \vlambda_{\rm c})$ the galaxy redshift posterior, and $\delta$ is a Dirac delta distribution of the galaxy sky position which can be treated as errorless. Equation~\ref {eq:pcat} assumes that the uncertainties in $w$ are negligible. The cosmological dependence enters because we do not know the true redshift, since the galaxy catalog contains observed redshift measurements $\tilde{z}_g$ and associated uncertainties $\tilde{\sigma}_{z,g}$. From these quantities, which we assume to be Gaussian, we construct the likelihoods, 
\begin{equation}\label{eq:pcat_gaussians}
     p(z \given \tilde{z}_g, \vlambda_\mathrm{c}) = \frac{\mathcal{N}(z; \tilde{z}_g, \tilde{\sigma}_{z,g}^2)\, p_\mathrm{bkg}(z, \mathcal{G} \given \vlambda_{\rm c})}{\int \mathcal{N}(z; \tilde{z}_g, \tilde{\sigma}_{z,g}^2)\, p_\mathrm{bkg}(z, \mathcal{G} \given \vlambda_{\rm c})\, \d{z}}.
\end{equation}
where $p_{\rm bkg}$ is our prior knowledge of the host background distribution. For instance, a typical assumption for a complete galaxy catalog is a uniform distribution in comoving volume, $p_{\rm bkg} \propto \d{V_{\rm c}}/\d{z}$. Alternatively, one could adopt a flat prior \citep[e.g.,][]{Gray2023}. We further discuss this quantity in the context of an incomplete catalog in Sect.~\ref{subsec:incompleteness}.

\subsection{Incompleteness}\label{subsec:incompleteness}
In GW cosmology, a catalog is considered ``complete'' when it contains all potential BBH host galaxies. In practice, galaxy observations are inevitably affected by selection effects that can depend on galaxy properties $\mathcal{G}$, limiting the completeness of the catalog. Therefore, Eq.~(\ref{eq:pcat}) has to be extended as follows \citep{Chen2018, Finke2021_DarkSirens}:
\begin{equation}
\begin{aligned}\label{eq:host_incomplete}
    p_{\rm gal}(z,\hat{\Omega},\mathcal{G}\given \vlambda_{\rm c}) &= f_\mathcal{R}\:\, p_{\rm cat}(z,\hat{\Omega},\mathcal{G}\mid \vlambda_{\rm c}) \\
    &\quad + \left(1-f_\mathcal{R}\right)\, p_{\rm miss}(z,\hat{\Omega},\mathcal{G}\mid \vlambda_{\rm c})\,,
\end{aligned}
\end{equation}
with 
\begin{equation}
    f_\mathcal{R} \equiv \int_0^{z_{\rm max}} P_{\rm comp} (z, \hat{\Omega}, \mathcal{G})\: p_{\rm bkg}(z, \mathcal{G} \given \vlambda_{\rm c})\: \d{z}\,,
\end{equation}
where $P_{\rm comp}$ is the completeness function, $p_{\rm bkg}$ is the background distribution of the potential hosts, and $f_\mathcal{R}$ is an overall completeness fraction that weights the contribution of the catalog term $p_{\rm cat}$ (Eq.~\ref{eq:pcat}) over the term representing the missing hosts $p_{\rm miss}$. Importantly, $f_\mathcal{R}$ must be calculated over a sufficiently large region up to $z_{\rm max}$ that includes all GW events that can be detected.

\begin{figure*}
    \centering
    \includegraphics[width=\linewidth]{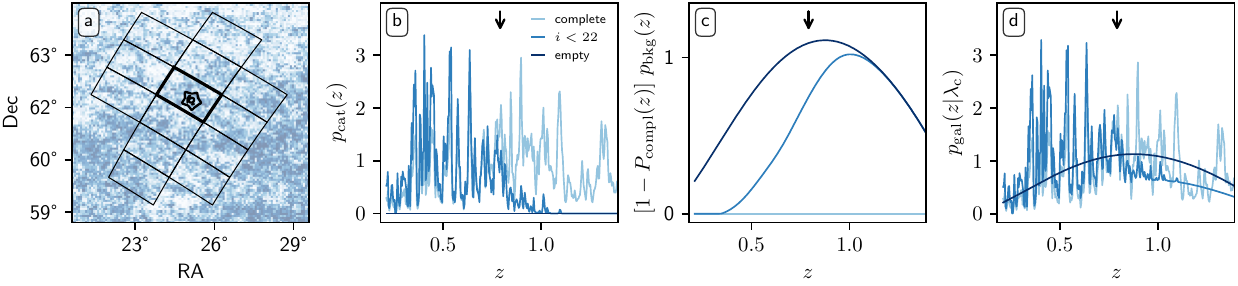}
    \caption{Effects of galaxy catalog incompleteness within the \texttt{CHIMERA} framework. (a) Sky localization and pixelization (represented with the grid) for a GW event from a stellar mass-weighted MICECATv2 catalog (the underlying blue points). The marker shows the host true position and pixel, from which the distributions in the subsequent panel are computed. (b) Catalog redshift distribution for complete (light blue), $i<22$ cut (blue), and empty catalogs (dark blue flat line). The arrow shows the true host redshift. (c) Completion terms accounting for missing galaxies; dark blue shows the background distribution used in standard siren constraints. (d) Resulting galaxy distribution terms, individually normalized for better comparison.}
    \label{fig:completeness_example}
\end{figure*}

The completeness function is defined in the domain $[0,1]$ and can be computed as
\begin{equation}\label{eq:completeness_function}
    P_{\rm comp}(z, \hat{\Omega}, \mathcal{G}) = \min\left(\frac{n_{\rm cat}(z, \mathcal{G})}{n_{\rm theo}(z, \mathcal{G})}, 1\right)\,,
\end{equation}
where $n_{\rm cat}(z)$ is the catalog number density and $n_{\rm theo}(z)$ is the  theoretical target number density, both depending on the properties of the host galaxies, $\mathcal{G}$.
Compared to the galaxy catalog term in Eq.~(\ref{eq:pcat}), $n_{\rm cat}(z)$ has to be computed over larger volumes to ensure adequate statistics and smooth out local inhomogeneities. We note that this approach, also used in \cite{Finke2021_DarkSirens}, differs from the one used in \texttt{gwcosmo} and \texttt{icarogw}, where the observed catalog distribution is instead modeled as the integral of a theoretical distribution down a limiting magnitude \citep{Gray2023, Mastrogiovanni2024_icarogw}. The key advantage of measuring the catalog density with a more physically motivated method is that it enables modeling completeness effects beyond simpler magnitude cuts, such as color-color selection or stellar mass cuts, more flexibly representing the selection functions of current galaxy surveys.

The distribution of galaxies can be described empirically using the parametric formulation introduced by \citet{Schechter1976}. It is convenient to express the theoretical number density of potential hosts as
\begin{equation}\label{eq:rho_theo}
    n_{\rm theo}(z, \mathcal{G}) = \int_{{\M}_{,\rm min}(z)}^{{\M}_{,\rm max}(z)} \Phi(\M,z)\, w(\M,z)\, \mathrm{d}\M\,,
\end{equation}
where $\Phi(\M,z)$ is the Schechter mass function, ${\M}_{,\rm min}(z)$ and ${\M}_{,\rm max}(z)$ are the minimum and maximum masses of galaxies hosting BBH mergers, and $w(\M,z)$ accounts for potential host weighting assumptions (see Sect.~\ref{subsec:gal} and Sect.~\ref{subsec:host}). The use of Schechter functions  is widespread in recent standard siren analyses, where galaxy luminosity or absolute magnitude is typically used as a weighting factor  \citep[e.g.,][]{Abbott2023_Cosmo, Mastrogiovanni2023, Turski2025}. Here, we instead weight by stellar mass directly to build a physically consistent framework tied directly to host galaxy properties. Moreover, value-added catalogs for stellar masses are expected to be released for current wide-field surveys such as DESI \citep{DESICollaboration2024} and \emph{Euclid} \citep{EuclidCollaboration2025}.
The background probability distribution of potential host galaxies, $p_{\rm bkg}$ can then be obtained by multiplying the theoretical galaxy density by the comoving volume element,
\begin{equation}\label{eq:pbkg}
   p_{\rm bkg}(z\given \vlambda_{\rm c},\mathcal{G}) \propto n_{\rm theo}(z, \mathcal{G})\,\frac{\d{V_{\rm c}(z)}}{\d{z}}\,.
\end{equation}
In general, this quantity depends on cosmological parameters, $\vlambda_{\rm c}$, but not on $H_0$, as $n_{\rm theo}$ scales as $H_0^3$ and $\d{V_{\rm c}}/\d{z}$ as $H_0^{-3}$.
Finally, under the assumption that missing galaxies are homogeneously distributed with respect to the background, we can write
\begin{equation}
    p_{\rm miss}(z, \hat{\Omega},\mathcal{G} \given \vlambda_{\rm c}) = \frac{1 - P_{\rm comp}(z, \hat{\Omega},\mathcal{G})}{1 - f_{\mathcal{R}}}\, p_{\rm bkg}(z,\mathcal{G} \given \vlambda_{\rm c})\,.
\end{equation}
This approach, known as ``homogeneous completion,'' differs from ``multiplicative completion,'' which assumes that missing galaxies trace the distribution of those present in the catalog \citep{Finke2021_DarkSirens}, and ``variance completion,'' which incorporates galaxy correlation lengths \citep{Dalang2024a, Leyde2024, Leyde2025}. In this work, we adopt the homogeneous completion method, which is self-consistent with the incompleteness effects simulated in our analysis (see Sect.~\ref{subsec:galobs}).
With all the necessary components in place, the galaxy term corrected for galaxy catalog incompleteness and host weighting becomes
\begin{equation}
\begin{aligned}\label{eq:pgal_final}
    &p_{\rm gal}(z,\hat{\Omega},\mathcal{G}\given \vlambda_{\rm c}) = \\
    &\quad \left[\int P_{\rm comp}(z,\hat{\Omega},\mathcal{G})\, p_{\rm bkg}(z,\mathcal{G}\given \vlambda_{\rm c})\,\d{z}\right]\, p_{\rm cat}(z,\hat{\Omega},\mathcal{G}\mid \vlambda_{\rm c}) \\
    &\quad + \left[1-P_{\rm comp}(z, \hat{\Omega},\mathcal{G})\right]\, p_{\rm bkg}(z,\hat{\Omega},\mathcal{G}\mid \vlambda_{\rm c})\,.
\end{aligned}
\end{equation}

\begin{figure*}
    \centering
    \includegraphics[width=1\linewidth]{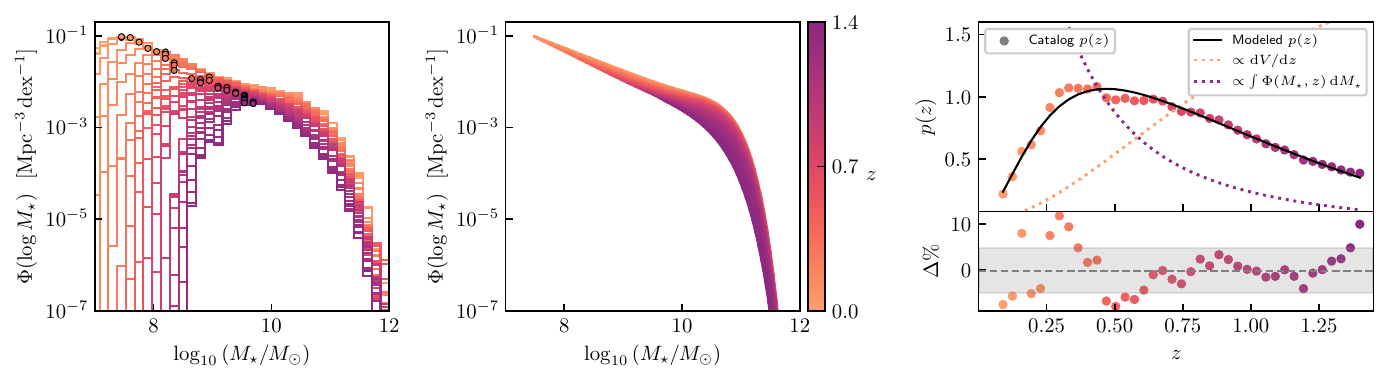}
    \caption{Stellar mass and redshift distribution of the parent galaxy catalog. \emph{Left panel:} Stellar mass function in different redshift bins measured from the original MICECATv2 galaxy catalog. The points indicate the peak number density at each redshift bin. \emph{Central panel:} Theoretical stellar mass function modeled as a double Schechter function evolving in redshift. \emph{Right panel:} Redshift distribution of the parent galaxy catalog. The black line shows the theoretical distribution obtained from the product of the cosmological volume element dominating at low-$z$ (dotted orange line) and the integral of the Schechter damping the distribution at high-$z$ (dotted purple line). The bottom panel shows the percentage difference, highlighting the median (gray line) and the 16-84 percentile range (gray band).
    }\label{fig:mice_Sch}
\end{figure*}

\subsection{Implementation in \texttt{CHIMERA}}\label{sec:implementation}
This section summarizes the implementation of the method in \texttt{CHIMERA} \citep{Borghi2024,Tagliazucchi2025}. The latest release (v2) presented in \cite{Tagliazucchi2025} is fully implemented in the \texttt{JAX} framework \citep{JAX}, enabling hyperparameter vectorization and GPU acceleration, and includes more efficient population functions and likelihood estimation algorithms. Together with this paper, we release an updated version of \texttt{CHIMERA} (v2.1) that enables proper treatment of galaxy catalog incompleteness and host weighting effects.\footnote{Available at \href{https://github.com/CosmoStatGW/CHIMERA}{https://github.com/CosmoStatGW/CHIMERA}}
\vspace*{-0.6em}
\paragraph{Likelihood.} The likelihood computation (Eq.~\ref{eq:hyperlike}) uses kernel density estimation (KDE) to model the GW data posterior. The KDEs are then integrated in equal-area \texttt{Healpix} pixels \citep{Gorski2005} that divide the sky area of the GW event (see Fig.~\ref{fig:completeness_example}a). \texttt{CHIMERA} employs three algorithms of increasing computational efficiency: (i) a full 3D kernel in (RA, Dec, $z$); (ii) a ``many-1d'' approach that marginalizes the 3D posterior using a 1D redshift KDE for each sky pixel; and (iii) a ``single-1d'' algorithm in which the GW information is collapsed onto a single sky pixel. The GW probability is then evaluated on a redshift grid where the KDE has support, considering all the hyperparameter space. In this paper, the (ii) approach is used. Although both approximate methods (ii) and (iii) show negligible differences from (i) for the datasets analyzed here \citep{Tagliazucchi2025}, we adopt the more conservative ``many-1d'' method.
\vspace*{-1.2em}
\paragraph{Catalog term.} To handle an efficient computation, the catalog term (Eq.~\ref{eq:pcat}) of a given pixelized GW dataset can be precomputed and saved to be loaded separately. This is crucial as there might be more than 1 million galaxies in a pixel. This quantity is $H_{0}$-independent as the numerator and denominator of the galaxy redshift posterior (Eq.~\ref{eq:pcat_gaussians}), scales in the same way as $H_{0}$, but must be re-computed for each host weighting scheme and galaxy catalog. Figure~\ref{fig:completeness_example}b shows an example for three catalogs: complete, i-band magnitude limited at $i<22$, and empty. The arrow indicates the redshift of the true host galaxy.
\vspace*{-1.2em}
\paragraph{Completeness.} The completeness function is computed in multiple $N_{\text{masks}}$ masks following the approach of \cite{Finke2021_DarkSirens}. Healpix pixels are grouped together using a k-means algorithm based on the number of galaxies inside each pixel. For each mask, the completeness function (Eq.~\ref{eq:completeness_function}) is evaluated by binning the observed galaxies into redshift bins and comparing their density to a theoretical background distribution. As for the catalog term, the completeness function can be saved to disk and reloaded to be later interpolated onto the specific redshift grid where the likelihood is evaluated. Note that the pixelization scheme used for the completeness can be coarser than that of the GW data. In this work, we adopt the default settings ($N_{\text{masks}} = 5$, \texttt{nside} = 32, and 100 redshift bins) that have been chosen to ensure adequate statistics within each mask.
\vspace*{-1.2em}
\paragraph{Background distribution.} The background galaxy distribution is obtained by integrating the Schechter mass function over the relevant host mass range with any applied host weighting (Eq.~\ref{eq:rho_theo}). This term is used in both the (pre)computation of the completeness function, $P_{\rm comp}$, and, crucially, is evaluated at each likelihood call and therefore must be defined efficiently, as it directly impacts the overall computational performance of the pipeline. In \texttt{CHIMERA}, this is optimized using just-in-time (JIT) compilation and vectorization across all redshift grids. An example illustrating the background galaxy distributions is shown in Figure~\ref{fig:completeness_example}c.
\vspace*{-1em}
\paragraph{Galaxy term.} The catalog and completeness correction terms are combined as in Eq.~(\ref{eq:pgal_final}). Figure~\ref{fig:completeness_example}d illustrates this interplay at different levels of completeness. For a complete catalog, all information comes from the catalog term, where galaxies are individually weighted according to the assumed hosting probabilities. On the other hand, in the absence of a galaxy catalog, all information comes from the completion term, and host weighting enters through the background distribution. In the intermediate case of a catalog complete down to a given magnitude, the relative contribution varies with redshift, showing a transition near $z \approx 1$ where the completeness begins to decline in the galaxy catalog adopted in this work (see Sect.~\ref{subsec:gal}).

\section{Catalogs}\label{sec:catalogs}

This section describes the generation of mock galaxy and GW catalogs. First, we build our parent galaxy catalog, which represents all galaxies in our synthetic universe and is described by a parametric, redshift-evolving stellar mass function (Sect.~\ref{subsec:gal}). From this parent catalog, we derive three sub-catalogs of potential BBH hosts by applying different host weighting schemes (Sect.~\ref{subsec:host}). For each host catalog, we generate mock GW events following a prescribed population model (Sect.~\ref{subsec:mockGW}) and simulate their detection with a Fisher matrix approach (Sect.~\ref{subsec:detGW}). Finally, we apply observational completeness cuts to the parent galaxy catalog to simulate realistic survey conditions (Sect.~\ref{subsec:galobs}). The parent galaxy catalog, either complete or with cuts applied, serves as the input catalog for the parameter estimation runs. Further details on this framework are provided in Appendix~\ref{app:mice_sch}.

\subsection{Parent galaxy catalog} \label{subsec:gal}
The mock galaxy catalog used in this work is obtained from the MICE Grand Challenge light-cone simulation (MICECATv2), which populates one-eighth of the sky ($5157~\mathrm{deg^2}$) and reproduces a Dark Energy Survey-type catalog completeness down to an observed magnitude of $i<24$ out to redshift $z=1.4$~\citep{Fosalba2015a,Fosalba2015b,Crocce2015,Carretero2015,Hoffmann2015}. The simulation assumes a flat $\Lambda$CDM cosmology with $H_0=70~\mathrm{km\,s^{-1}\,Mpc^{-1}}$, $\Omega_{m,0}=0.25$, and $\Omega_{\Lambda,0}=0.75$.

We select a subsample of MICECATv2 so that the redshift distribution can be described with a parametric function and efficiently computed at each likelihood call of \texttt{CHIMERA}, while preserving realistic galaxy clustering. We first compute the stellar mass distribution of the original catalog, defining the minimum mass as the mass corresponding to the peak density at each redshift bin (Fig.~\ref{fig:mice_Sch}, left panel). We then model this distribution with a redshift-evolving double Schechter function (Fig.~\ref{fig:mice_Sch}, central panel), where the evolution is driven by the minimum mass. Lastly, we subsample the original catalog so that our model can describe the background galaxy distribution of the catalog with a typical percentage difference below $5\,\%$ (Fig.~\ref{fig:mice_Sch}, right panel). To compute the theoretical number density (Eq.~\ref{eq:rho_theo}) the Schechter function is integrated from a redshift-evolving minimum mass to $\log_{10}(M_\star/M_\odot) \approx 12$. Further details are given in Appendix~\ref{app:mice_sch}. The resulting catalog, identified as the 
parent galaxy catalog throughout this paper, contains approximately 335 million galaxies (about $67\%$ of MICECATv2) and is more than 200 times larger than the high-mass subsample of MICECATv2 galaxies used in \citetalias{Borghi2024}.

\subsection{Potential host galaxies} \label{subsec:host}
The connection between BBH and host galaxy properties is a subject of ongoing research. On theoretical grounds, population synthesis studies suggest that low-$z$ BBH mergers preferentially occur in galaxies with higher stellar mass and more metal-rich \citep[e.g.,][]{Artale2020, Santoliquido2022}. On the observational side, GW population studies are only beginning to explore these dependencies \citep[e.g.,][]{Vijaykumar2024}.

\begin{figure}
    \centering
    \includegraphics[width=.95\hsize]{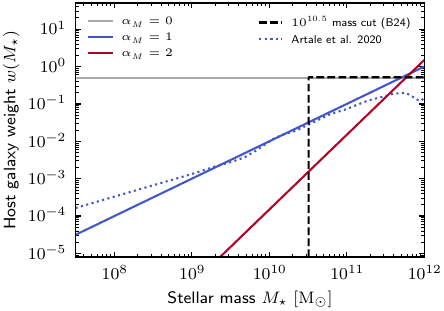}
    \caption{Host galaxy weighting schemes considered in this work: constant (gray), linear (blue), and quadratic (red) mass weighting.  For comparison, we show the $z=0.1$ trend from \citet{Artale2020} (dotted blue line) and the cut used in previous work \citep[solid line, as in][]{Borghi2024}.}
    \label{fig:w}
\end{figure}

In this study, we assign the probability that each galaxy hosts a BBH merger as a function of its stellar mass without applying mass cuts. We consider three weighting schemes:
\begin{equation}\label{eq:weight_schemes}
    w \propto M_\star^{\alpha_M}, \quad \text{where} \quad 
    \alpha_M = 0,\,1,\,2.
\end{equation}
These cases are compared in Fig.~\ref{fig:w} alongside the weighting adopted in \citetalias{Borghi2024} and the low-$z$ trend from the population synthesis study of \citet{Artale2020}.
Despite the uniform ($\alpha_M=0$) weighting being maximally agnostic, it unrealistically assigns equal hosting probability to all galaxies, regardless of their stellar mass, across a range spanning several orders of magnitude. Applying a stellar-mass cut is a practical attempt to mitigate this issue. The $\alpha_M=1$ weighting is more in line with theoretical predictions in the local Universe, where most of the BBH events are observed (see Sect.~\ref{subsec:detGW}). Finally, the $\alpha_M=2$ weighting allows us to explore the case of an even stronger dependence on stellar mass.

We remark that the parent catalog is complete above a given minimum stellar mass that evolves with redshift (Fig.~\ref{fig:mice_Sch}). In the likelihood, we integrate the Schechter function over this mass range to ensure statistical consistency. This allows us to probe the regime in which the galaxy density is highest and therefore least informative for the dark siren method. In practice, this occurs for $\alpha_M=0$ at low $z$, where most of the events are located. However, since low-mass galaxies are increasingly missed at higher redshift, caution would be required if extending this analysis to constrain $H(z)$ at higher $z$ with the $\alpha_M=0$ catalog (or even more so with $\alpha_M<0$), as the galaxy density of this parent sample may not necessarily reflect the true underlying population. This issue does not affect mocks generated with $\alpha_M>0$, since the weights preferentially select high-mass galaxies where the fidelity of the parent sample is highest.

\subsection{Mock GW catalogs} \label{subsec:mockGW}
For each weighting scheme, we populate the corresponding host galaxy catalog with BBH mergers, generating three mock GW catalogs hereafter labeled as \texttt{mock0}, \texttt{mock1}, \texttt{mock2} for $\alpha_M = 0, 1, 2$ respectively. We assume that the overall BBH merger rate follows the cosmic star formation rate density from \citet{Madau2014}:
\begin{equation}\label{eq:rate_MD}
    \psi(z ; \vlambda_{\rm z}) \propto \frac{(1+z)^{\gamma}}{1 + \left (  \frac{1+z}{1+z_\mathrm{p}}   \right )^{\gamma + \kappa}},
\end{equation}
where $\psi(z) \propto (1+z)^{\gamma}$ at low $z$, peaks at $z_{\rm p}$, and then declines as $\psi(z) \propto (1+z)^{-\kappa}$. The redshifts are then converted to luminosity distances by assuming the cosmology of MICECATv2. For the mass distribution, we adopt the phenomenological model ``PowerLaw+Peak'' \citep[PLP,][]{Talbot2018}, which is the preferred model up to LVK GWTC-3 \citep{Abbott2023_Population}. Equation~(\ref{eq:pop}) is then factorized as
\begin{align}
    p(m_1,m_2|\vlambda_{\rm m}) = p(m_1|\vlambda_{\rm m})\, p(m_2|m_1, \vlambda_{\rm m}),
\end{align}
where the primary mass $p(m_1|\vlambda_{\rm m})$ is modeled as a superposition of a power-law with slope $\alpha$ in the domain $m_1\in [m_{\rm low},m_{\rm high}]$ and a Gaussian centered at $\mu_{\rm g}$ with width $\sigma_{\rm g}$ and mixing fraction $\lambda_{\rm g}$. The low-mass boundary is smoothed by a factor set by $\delta_m$. The secondary mass is modeled as a power-law with index $\beta$ in the domain $m_2 \in [m_{\rm low}, m_1]$.\\

In summary, our GW population is described by these twelve cosmological and astrophysical hyperparameters:
\begin{equation}\label{eq:allparams}
\begin{split}
        \vlambda_{\rm c} &= \{H_0\} \\
        \vlambda_{\rm z} &= \{\gamma, k, z_{\rm p} \} \\
        \vlambda_{\rm m} &= \{\alpha, \beta, \delta_m, m_{\rm low}, m_{\rm high}, \mu_{\rm g}, \sigma_{\rm g}, \lambda_{\rm g} \}.
\end{split}
\end{equation}
The fiducial values and prior ranges chosen for this work are the same as those in \citetalias{Borghi2024} (see their table 1), ensuring that the forecasts can be directly compared in terms of the GW population. In particular, we fix $\Omega_{m,0}$ to the value adopted in MICEcatv2.

\begin{figure*}
    \centering
    \includegraphics[width=1\linewidth]{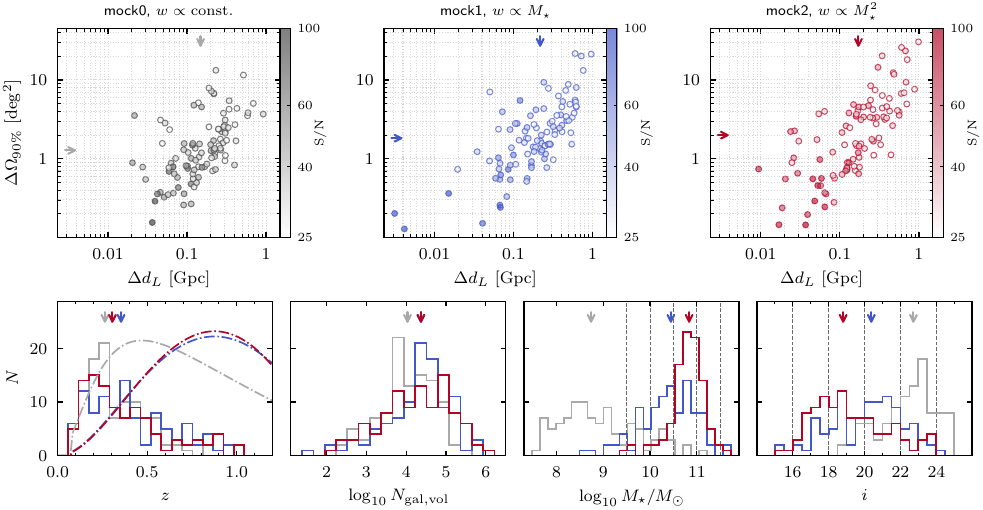}
    \caption{Main properties of 100 O5-like BBH events with $\mathrm{S/N} > 25$ drawn from MICECATv2 with three host weighting schemes: unweighted (\texttt{mock0}, gray), weighted by stellar mass (\texttt{mock1}, blue), and weighted by stellar mass squared (\texttt{mock2}, red). \emph{Top panels:} Distribution in the sky localization ($90\%$ C.L.) versus luminosity-distance uncertainty plane color-coded by the network $\mathrm{S/N}$. \emph{Bottom panels:} Distributions of source redshift, number of galaxies within the localization volume ($N_{\mathrm{gal,vol}}$), host galaxy stellar mass, and $i$-band magnitude for the various samples; the arrows indicate the median values. The dot-dashed lines show the background redshift distribution of potential BBH host galaxies used to generate GW events. The vertical dashed line shows the mass and magnitude cuts that are applied in this work to study incompleteness effects.}\label{fig:mock_properties}
\end{figure*}

\subsection{Detected GW events} \label{subsec:detGW}
To simulate detections, we use \texttt{GWFAST} \citep{Iacovelli2022,Iacovelli2022_GWFAST}. We assume quasi-circular non-precessing BBH systems so that the following detector frame parameters characterize each waveform:
\begin{equation}
    \boldsymbol{\theta}^{\rm det} = \{\mathcal{M}, \eta, d_L, \theta, \phi, \iota, \chi^z_1, \chi^z_2, \psi, t_{\rm c}, \Phi_{\rm c} \},
\end{equation}
where $\mathcal{M}$ is the detector-frame chirp mass, $\eta$ is the symmetric mass ratio, $d_L$ is the luminosity distance, $\theta = \pi/2 - \mathrm{Dec}$ and $\phi = \mathrm{RA}$ are the sky position angles, $\iota$ refers to the inclination angle of the binary's orbital angular momentum with respect to the line of sight, $\chi_{1/2, z}$ are the dimensionless spin parameters along the direction of the orbital angular momentum, $\psi$ is the polarization angle, $t_c$ is the coalescence time, and $\Phi_c$ is the phase at coalescence. The first five parameters are drawn from mock GW catalogs presented in Sect.~\ref{subsec:mockGW}, while the remaining parameters follow standard priors: inclination angle $\cos\iota{\sim} \mathcal{U}(0,\pi)$; spin components $\chi_{1,2}^z{\sim}\mathcal{U}(-1,1)$; polarization angle $\psi{\sim}\mathcal{U}(0,\pi)$, coalescence time $t_{\rm c}{\sim}\mathcal{U}(0,1)$ (in units of fraction of a day), and coalescence phase $\Phi_{\rm c}{\sim}\mathcal{U}(0,2\pi)$.
For each source, we simulate GW emission and compute the network signal-to-noise ratio ($\mathrm{S/N}$) using the \texttt{IMRPhenomHM} \citep{London2018,Kalaghatgi2020} waveform approximant. In this work, we consider an O5-like network configuration,\footnote{The amplitude spectral densities can be found at~\url{https://dcc.ligo.org/LIGO-T2000012/public}. In particular, we use \texttt{AplusDesign} for the three LIGO detectors, \texttt{avirgo\_O5low\_NEW} for Virgo, and \texttt{kagra\_80Mpc} for KAGRA.} which includes the two LIGO \citep{LIGOScientificCollaboration2015}, Virgo \citep{Acernese2015}, and KAGRA \citep{Aso2013} instruments, as well as a LIGO detector located in India~\citep{Abbott2016_Prospects}. We assume a minimum frequency of $10~\mathrm{Hz}$ and a duty cycle of $100\%$.

We select a subsample of 100 BBHs with a network signal-to-noise ratio, $\mathrm{S/N}>25$. This cut, chosen to be consistent with \citetalias{Borghi2024}, is designed to yield the 100 best BBH events over approximately one year of observation. In general, higher S/N events have tighter sky localization and fewer potential host galaxies, resulting in stronger $H_0$ constraints in the dark siren regime. In contrast, pure spectral siren analyses (empty catalog limit), included here for comparison, would benefit from including lower S/N events to better resolve features in the mass distribution \citep[see e.g.,][]{Abbott2023_Population}.

Figure~\ref{fig:mock_properties} presents the main properties of the 100 BBH events generated for each of the three host weighting schemes. The top panels show the $90\,\%$ C.L. sky localization area against the luminosity distance uncertainty, color-coded by $\mathrm{S/N}$. The bottom panels show the redshift distribution, galaxy counts within the localization volume ($N_{\mathrm{gal,vol}}$), and host galaxy properties (stellar mass $\M$ and $i$-band magnitude). Note that $N_{\mathrm{gal,vol}}$ assumes a flat prior on $H_0 \in [60, 80]~\mathrm{km\,s^{-1}\,Mpc^{-1}}$.
The observed GW events are located at $z\lesssim 1$, with sky localization areas ranging from $0.1$ to a few tens of square degrees. The values of $N_{\mathrm{gal,vol}}$, a key indicator of how strongly each GW event can constrain $H_0$ in the dark siren regime, range from $\mathcal{O}(10^2)$ to $\mathcal{O}(10^{5})$, with the maximum set by the finite redshift coverage of the current catalog. These values are significantly larger than those in \citetalias{Borghi2024}, where the best-localized events contained only a few galaxies and could therefore be treated as ``pseudo-bright'' sirens. Differences among the three mocks reflect both the background host galaxy distribution (see Fig.~\ref{fig:mock_properties}, bottom left panel) and the specific realization of the sampled GW events. Due to the lower redshift distribution of their hosts, \texttt{mock0} events are observed with slightly higher $\mathrm{S/N}$ and better sky localization, with median values of $34.1$ and $1.3~\mathrm{deg^2}$, respectively, compared to \texttt{mock1}  ($32.5$ and $1.8~\mathrm{deg^2}$) and \texttt{mock2} ($31.6$ and $2.0~\mathrm{deg^2}$). As a result, the median $N_{\mathrm{gal,vol}}$ differs between the mocks: $\approx 6\,500$ for \texttt{mock0} and $\approx 12\,000$ for both \texttt{mock1} and \texttt{mock2}.

We compute selection effects by generating a large set of injections with \texttt{GWFAST} using the same $\mathrm{S/N}>25$ threshold applied to the construction of the GW catalog. We verify that these events span the full range of detectable luminosity distances and binary masses. The full population includes $2\times10^7$ generated sources, resulting in $1\times10^6$ detections at $\text{S/N}>25$ in the O5-like scenario considered in this work. This dataset is then used to estimate the selection effects following Eq.~(\ref{eq:sel_func_est}).

\subsection{Observed galaxy catalogs} \label{subsec:galobs}
Galaxy catalogs are generally constructed to achieve a well-defined completeness threshold, typically set by apparent magnitude or stellar mass limits. Magnitude-complete surveys maintain uniform flux thresholds across the survey footprint. The Galactic plane, which covers approximately one third of the full sky, is typically surveyed at shallower magnitudes or excluded from deeper surveys because of severe dust extinction, which limits the detection of fainter galaxies. Some well-known examples include SDSS \citep{York2000}, GAMA \citep{Driver2011}, DESI \citep{DESICollaboration2024}, zCOSMOS \citep{Lilly2007}, and VIPERS \citep{Guzzo2014}. Mass-complete catalogs are built from subsamples of galaxies that are selected to ensure homogeneity in terms of intrinsic physical properties. As a result, these catalogs are better suited to study the evolution of galaxies throughout cosmic time \citep[e.g.,][]{Pozzetti2010, Weaver2023}. 

In this work, we consider the case of a spectroscopic galaxy catalog with uncertainties on the galaxy redshifts of $\sigma_z = 0.001(1+z)$. We then generate catalogs including incompleteness effects by removing galaxies according to the following mass and magnitude cuts:
\begin{equation}\label{eq:completeness_cuts}
\begin{aligned}
    \log_{10}\M > \log_{10}\Mcut &\in \{9.5, 10, 10.5, 11, 11.5\} \\
    i < i_{\rm cut} &\in \{24, 22, 20, 18, 16\}
\end{aligned}
\end{equation}
We also examine two limiting cases: a complete catalog (no selection cuts) representing ideal survey conditions, and an empty catalog (spectral siren analysis only) representing the worst-case scenario where no potential host galaxies are observed.

\begin{figure*}
    \centering
    \includegraphics[width=\linewidth]{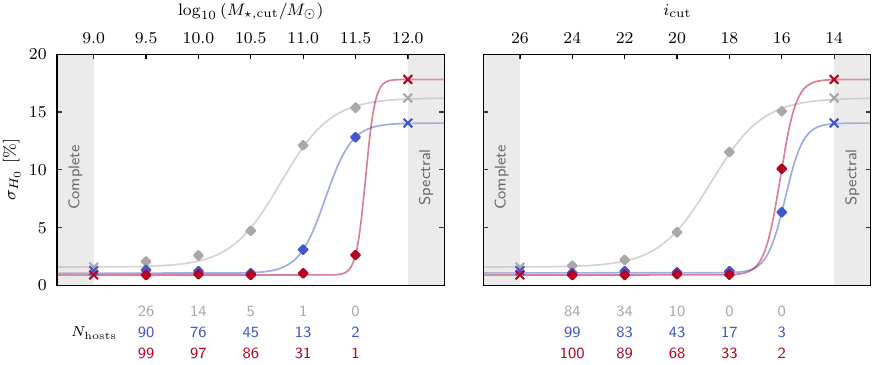}
    \caption{Effects of galaxy survey incompleteness on $H_0$ constraints for different stellar mass (left panel) and $i$-band magnitude (right panel) completeness cuts across three host weighting schemes: unweighted (\texttt{mock0}, gray), weighted by stellar mass (\texttt{mock1}, blue), and weighted by stellar mass squared (\texttt{mock2}, red). The values annotated below each panel indicate the number of host galaxies retained in the catalog after applying each cut.}
    \label{fig:H0_completeness}
\end{figure*}

\section{Results}\label{sec:results}

This section presents the results of the full standard siren analysis for all 36 configurations, combining three host galaxy weighting schemes (Eq.~\ref{eq:weight_schemes}) with twelve catalog completeness cuts (Eq.~\ref{eq:completeness_cuts}, including the complete and empty cases). 

The posterior distribution is sampled with the affine-invariant MCMC sampler \texttt{emcee} \citep{emcee}, adopting wide uniform priors. We ensure convergence by requiring that the number of samples is at least $50\times$ the integrated autocorrelation time for each hyperparameter. The entire analysis, enabled by the enhanced version of \texttt{CHIMERA} \citep{Tagliazucchi2025}, took about $50\,\mathrm{k}$ CPU hours on the LEONARDO supercomputer hosted by CINECA \citep{Turisini2024}. The results are quoted as median values with symmetric $68.3\%$ credible intervals. 

Figure~\ref{fig:H0_completeness} shows the marginalized constraints on $H_0$ in all the completeness and weighting configurations. The full table of results can be found in Appendix~\ref{app:res_table}.
We obtain unbiased estimates of $H_0$ within $1\,\sigma$ across all configurations, even at high levels of incompleteness. The remaining hyperparameters are all recovered within $2\,\sigma$. This demonstrates the robustness of both the statistical methodology and the framework used to generate mock catalogs. 
In the following sections, we present the results for the complete catalog, empty catalog, and intermediate completeness configurations. We then examine the correlations between $H_0$ and the hyperparameters describing the astrophysical population. Finally, we assess potential biases that may arise from wrong assumptions on the host galaxy weighting.

\subsection{Complete galaxy catalog} \label{sec:results:complete}
When the galaxy catalog contains all potential hosts, we obtain percent-level constraints on $H_0$. The precision improves when host weighting favors more massive galaxies, with $\sigma_{H_0}=1.6\%,\,1.3\%,0.9\%$ for \texttt{mock0}, \texttt{mock1}, \texttt{mock2}, respectively. The main reason for this improvement is the decrease in the effective number of galaxies within the GW localization volume, $N_{\rm gal, vol}$, that significantly contribute to the catalog term once the weighting is applied. We verify that the trend is independent of the specific realization of GW events by performing one-dimensional analyses on the $H_0$ posterior, keeping all the other hyperparameters fixed to their fiducial values. This is further supported by the fact that \texttt{mock0}, despite having lower $N_{\rm gal, vol} $ values (see Fig.~\ref{fig:mock_properties}, bottom panels), provides the weakest constraints.

These results are remarkable given that typical $N_{\rm gal, vol}$ values are $200\times$ larger than in \citetalias{Borghi2024}. Even with this increased galaxy density, constraints degrade by only a factor of two in the least informative case. We attribute this to the fact that, even when the analysis is extended to more realistic and denser catalogs, the redshift uncertainties provided by a spectroscopic galaxy catalog remain sufficiently small to preserve the information content of the catalog term (Eq.~\ref{eq:pcat}). Consequently, through the combined analysis of multiple events, the true value of $H_0$ can still be statistically recovered with high precision.

\subsection{Empty galaxy catalog} \label{sec:results:empty}
When no galaxies are included, cosmological constraints are obtained with a pure spectral siren approach, which relies on assumptions on the source-frame BBH mass distribution to break the mass-redshift degeneracy. We find $\sigma_{H_0}=16\%,\,14\%,18\%$ for \texttt{mock0}, \texttt{mock1}, \texttt{mock2}, respectively. These results are consistent with forecasts based on similar methodologies and BBH population assumptions \citep{Mancarella2022,Leyde2022,Borghi2024}. Unlike the complete catalog case, the trend with the weighting exponent $\alpha_M$ is no longer present here, as the constraints depend more strongly on the specific realization of the GW events, in particular how well the relevant features of the mass distribution are mapped. These spectral siren results establish the limiting case for studying how cosmological constraints degrade as galaxy catalog completeness decreases.

\subsection[Modeling incompleteness effects on H0]{Modeling incompleteness effects on H\textsubscript{0}}\label{sec:results:incompleteness}
To describe how galaxy catalog incompleteness affects the precision of $H_0$ constraints, we model the evolution of the percentage uncertainty, $\sigma_{H_0}$, using a sigmoid function that smoothly connects the case of a complete catalog ($\sigma_{H_0}^{\rm comp}$) to the empty catalog/spectral siren limit ($\sigma_{H_0}^{\rm spec}$):
\begin{equation}
    \sigma_{H_0}(x) = \frac{\sigma_{H_0}^{\rm spec} - \sigma_{H_0}^{\rm comp}}{1 + \exp[-k(x - x_{\mathrm{th}})]} + \sigma_{H_0}^{\rm comp},
\end{equation}
where $x$ is the chosen completeness parameter (such as stellar mass or magnitude), $x_{\mathrm{th}}$ is the threshold at which the uncertainty transitions most rapidly between the two regimes, and $k$ sets the steepness of the transition.

For mass-completeness cuts, we find $x_{\rm th}\equiv \log_{10}M_{\rm th}=10.8,\ 11.2,\ 11.6$ and $k = 4.2,\ 11.1,\ 22.4$ for \texttt{mock0}, \texttt{mock1}, and \texttt{mock2}, respectively. As $\alpha_M$ increases, the threshold shifts toward higher stellar masses, indicating that missing less massive (less probable) hosts ultimately has a negligible impact on the $H_0$ constraints. Notably, with mass and mass-squared weighting, percent-level precision is still achievable as long as the catalog is complete at $\log_{10} (\M/M_\odot) > 11$. The annotations in Fig.~\ref{fig:H0_completeness} show the number of true host galaxies remaining after each completeness cut. Percent-level constraints are obtained when at least about $20$ events retain their hosts after completeness cuts or, equivalently, when the catalog completeness is high enough to keep a sufficient fraction of true host galaxies in the catalog. Interestingly, there is one configuration (\texttt{mock2} with $\log_{10}(M_{\rm cut}/M_\odot)=11.5$) for which this is possible thanks to a single lucky event. We underscore the importance of spectroscopic follow-ups to target the localization region of these best-constraining ``golden'' events with future galaxy surveys.

\begin{figure}[t]
    \centering
    \includegraphics[width=\linewidth]{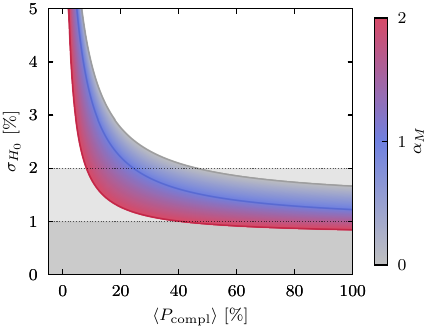}
    \caption{Percentage uncertainty on marginalized $H_0$ constraints as a function of the average completeness and host mass weighting exponent $\alpha_M$ obtained from a full standard siren analysis of 100 O5-like BBH events. }
    \label{fig:H0_completeness_smooth}
\end{figure}

\begin{figure}[t]
    \centering
    \includegraphics[width=1\linewidth]{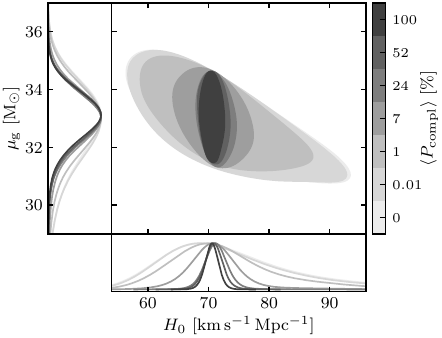}
    \caption{Constraints on the $H_0-\mu_{\rm g}$ plane obtained with 100 O5-like BBH events from the unweighted (\texttt{mock 0}) dataset at varying galaxy catalog average completeness.}
    \label{fig:contours_mcmc}
\end{figure}

For magnitude-completeness cuts, we find $x_{\rm th}\equiv i_{\mathrm{th}} = 18.7, 15.9, 16.0$ and $k = 1.0, 3.0, 3.3$ for \texttt{mock0}, \texttt{mock1}, and \texttt{mock2}, respectively. We do not observe a clear trend of the threshold with $\alpha_M$, as the difference between \texttt{the mock1} and \texttt{mock2} results is mainly driven by the specific realization of the GW events analyzed. This is also expected given that the distributions of $i$-band magnitudes of the true hosts are quite similar in the two mocks (see Fig.~\ref{fig:mock_properties} lower right). In both \texttt{mock1} and \texttt{mock2}, percent-level constraints can be obtained with catalogs containing all potential BBH hosts down to magnitudes $i<17$. However, we emphasize that realistic catalogs are characterized by specific and more complex selection functions; therefore, caution should be taken when extrapolating these results to the observed surveys. 

To summarize our results, we compute the comoving volume-averaged completeness within the GW detector horizon,
\begin{equation}\label{eq:avg_Pcompl}
    \langle P_{\rm compl}\rangle \equiv \frac{1}{V_{\rm c}(z_{\rm hor})}\int_0^{z_{\rm hor}} P_{\rm compl}(z)\,\d{V_{\rm c}}\quad\text{with}\quad z_{\rm hor} = 1.4\,,
\end{equation}
where we recall that $P_{\rm compl}$ refers to potential BBH host galaxies rather than the full galaxy population and, being integrated over the GW horizon, is non-local.
While a more local definition would better capture the completeness of the most informative events and correlate more directly with the $H_0$ constraints, we adopt this integrated metric to ensure a consistent comparison across all the different configurations.

Figure~\ref{fig:H0_completeness_smooth} shows the trends of $\sigma_{H_0}$ at varying $\langle P_{\rm compl}\rangle$, smoothed with a Gaussian kernel and interpolated as a function of $\alpha_M$. In the worst-case scenario of uniform weighting, achieving a $2\%$ constraint on $H_0$ requires a spectroscopic catalog with at least $50\,\%$ completeness within the GW detector horizon. In the more realistic case of mass weighting, the same completeness level would provide a $1.5\%$ constraint, while achieving $1\%$ precision is only possible if BBH hosting probability scales with the stellar mass squared.

\subsection{Correlation with astrophysical parameters}\label{sec:results:correlations}
The study of correlations between $H_0$ and astrophysical hyperparameters across the different configurations provides insights into how strongly cosmological constraints depend on assumptions about the BBH mass distribution. When the catalog term is not informative, a strong anticorrelation typically appears between $H_0$ and the mass scale parameter $\mu_{\rm g}$. This happens because higher $H_0$ values shift sources to higher $z$ at fixed luminosity distance, requiring smaller source-frame masses to match the observations. This behavior has been observed in real data \citep[e.g.,][]{Abbott2023_Cosmo, Mali2025, GWTC4_Cosmo}, in mock analyses when using photometric galaxy catalogs \citep{Borghi2024}, and is clearly evident in this work as the completeness decreases.

We generally find a strong anti-correlation between $H_0$ and $\mu_{\rm g}$ at average completeness levels below 10\%, with Spearman correlation coefficients computed from MCMC chains reaching values of $\rho \approx -0.7$. When the catalog completeness is greater than 50\%, these correlations weaken substantially. Percent-level constraints on $H_0$ are obtained only when $|\rho|<0.15$, at which point the anti-correlation is broken for all three mocks. Figure~\ref{fig:contours_mcmc} illustrates the case of \texttt{mock 0}, where the anti-correlation is broken at completeness levels above $25\,\%$, which correspond, in our mock catalogs, to a stellar mass completeness cut of $\log_{10}(\M/M_\odot)>10$. We also observe weaker, less significant anti-correlations between $H_0$ and the edges of the mass distribution $m_{\rm high}$ and $m_{\rm low}$.

\subsection{Mismodeling hosting weights}\label{sec:results:syst}
We investigate potential biases introduced by incorrect host galaxy weighting assumptions by re-analyzing the GW events from \texttt{mock1} using incorrect weighting schemes during parameter estimation (i.e., $\alpha_M=0$ and $\alpha_M=2$), while keeping the parent catalog and all other assumptions fixed. These weighting choices propagate through the likelihood by affecting individual galaxy weights (Eq.~\ref{eq:pcat}), the completeness function (Eq.~\ref{eq:completeness_function}), and the background distribution (Eq.~\ref{eq:pbkg}). Further details and supporting figures are provided in Appendix~\ref{app:syst}.

For complete spectroscopic catalogs, incorrect weighting introduces negligible systematic bias ($|\Delta H_0| < 1$ km s$^{-1}$ Mpc$^{-1}$), with mass-squared weighting surprisingly producing tighter constraints. At moderate average completeness levels (${\approx}20\%$), a mass weighting stronger than the true one ($\alpha_M=2$) still yields unbiased results, while uniform weighting ($\alpha_M=0$) introduces biases of ${\approx}10$ km s$^{-1}$ Mpc$^{-1}$. When catalogs become highly incomplete or empty (spectral siren regime), both incorrect weighting assumptions produce biases, exceeding $1\,\sigma$ for $\alpha_M = 0$ and $2\,\sigma$ for $\alpha_M = 2$. This demonstrates the importance of accurately modeling the host population when galaxy catalog information becomes sparse.

Our results are in agreement with previous studies of host galaxy weighting systematics. In particular, \cite{Hanselman2025} and \cite{Alfradique2025} also find unbiased $H_0$ constraints in the case of a complete catalog with stellar-mass mismodeled by equal weighting. In this work, we confirm the existence of a regime where host weight mismodeling does not affect $H_0$ constraints also at higher level of incompleteness. 
The robustness of these constraints is likely due to the assumption of very informative catalog terms with spectroscopic redshift uncertainties. 
However, assumptions on the GW data can also play a role. Dedicated mock data challenges will be crucial to compare analysis frameworks and better assess these effects.

\section{Conclusions}\label{sec:conclusions}

In this work, we study the cosmological potential of standard sirens, focusing on the impact of galaxy catalog incompleteness and host galaxy weighting on $H_0$ constraints. Below, we summarize the main results of the paper.

\begin{enumerate}
    \item We present a self-consistent framework to study galaxy catalog incompleteness and host weighting effects in standard siren analyses and implement it in the publicly available \texttt{CHIMERA} pipeline (v2.1).\footnote{Available at \href{https://github.com/CosmoStatGW/CHIMERA}{https://github.com/CosmoStatGW/CHIMERA}} The updated version of the code is designed to be more modular and efficient for large galaxy catalogs.

    \item We build our parent sample from the MICECATv2 mock galaxy catalog and describe it with a parametric stellar mass distribution. We then generate three realizations of 100 BBH events at $\mathrm{S/N} > 25$ detectable by a LIGO-Virgo-KAGRA O5-like network over about one year of observation. Each realization is based on a different host galaxy weighting scheme: unweighted, weighted by stellar mass, and weighted by stellar mass squared. We explore standard siren constraints achievable with spectroscopic catalogs, simulating incompleteness through stellar mass or magnitude cuts.

    \item We obtain percent-level constraints on $H_0$ even at moderate incompleteness. In the best-case scenario of a complete spectroscopic catalog, precision reaches 1.6\%, 1.3\%, and 0.9\% for no weighting, mass weighting, and mass-squared weighting, respectively. Remarkably, $2\%$ precision is achievable even when the completeness of potential hosts, averaged within the GW horizon, reaches $50\%$. Achieving $1\%$ precision in the assumed configurations is only possible if BBH hosting probability scales with the stellar mass squared.
    
    \item We characterize how $H_0$ constraints degrade with decreasing catalog completeness using a sigmoid function. We find that the completeness threshold and the steepness of the sigmoid transition increase for higher mass weighting schemes, demonstrating quantitatively that missing progressively more low-mass host galaxies has negligible impact on the cosmological constraints.

    \item We study the correlations between $H_0$ and the astrophysical population parameters. A strong anti-correlation between $H_0$ and the BBH mass scale parameter $\mu_{\rm g}$ (with Spearman coefficients $\rho \approx -0.7$) is present at average completeness levels below ${\approx} 10\%$. This correlation weakens substantially above $50\,\%$ completeness. Percent-level constraints on $H_0$ are achieved when $|\rho| < 0.15$, demonstrating the transition from the spectral to the dark siren regime.

    \item We assess systematic effects from mismodeling host galaxy weighting. With complete spectroscopic catalogs, incorrect weighting does not introduce significant $H_0$ biases. At moderate average completeness (${\approx}20\%$), stronger-than-true mass weighting yields unbiased results, while weaker weighting introduces $10~\mathrm{km\,s^{-1}\,Mpc^{-1}}$ biases. For highly incomplete or empty catalogs (spectral siren regime), mismodeling the weighting in either direction introduces significant biases, highlighting the importance of accurately modeling the host population when galaxy catalog information becomes sparse.
    
\end{enumerate}

These results are based on some simplifying assumptions. First, we assume that galaxy properties are measured with negligible uncertainty. Second, we assume that we know the true underlying galaxy distribution. Third, we simulate catalog incompleteness with simple stellar mass and magnitude cuts, while actual surveys follow more complex selection functions as they might target different galaxy populations and be subject to different selection effects. In future work, we plan to test the impact of these assumptions by applying realistic survey conditions tailored to ongoing and upcoming large-scale surveys. We also plan to assess the impact of photometric redshift measurements on our results, with particular focus on the trade-off between catalog completeness and the precision of redshift measurements. Together, these steps will be crucial to push the cosmological constraints to higher $z$ and study the expansion history $H(z)$.
Finally, our study assumes a duty cycle of 100\% at nominal detector sensitivity for all instruments in LVK O5-like configuration, providing a simple benchmark for expected yearly O5 performance. Under more realistic observing conditions, achieving the same precision on $H_0$ would require proportionally longer observation times.
Despite these limitations, our results highlight the crucial importance of spectroscopic galaxy surveys for standard siren cosmology. Moreover, our framework demonstrates that even moderately incomplete spectroscopic catalogs can yield competitive cosmological constraints, even when hosting probabilities are not properly modeled.

To put our results in context, current constraints from GWTC-4.0 provide ${\approx}20\%$ uncertainties on $H_0$ from 141 CBC dark sirens, using a galaxy catalog where only about one quarter of the galaxies have spectroscopic redshifts \citep{GWTC4_Cosmo}. Our forecast of $1{-}2\%$ precision with 100 O5-like BBH events and a spectroscopic catalog confirms and extends the findings of our previous work \citepalias{Borghi2024}, which achieved percent-level constraints using a complete sample of massive hosts with $\log_{10}(\M/M_\odot) > 10.5$. Here we validate those results across the full galaxy-mass range and present a comprehensive end-to-end analysis of catalog incompleteness and host weighting effects on dark-siren constraints. Our approach employs realistic mock catalogs, marginalization on the astrophysical population parameters, and hierarchical inference consistent with current LVK cosmology analyses, extending earlier explorations \citep{Gray2020} and going beyond more simplified analytical models used in recent studies \citep{Perna2025, Hanselman2025, Alfradique2025, Cross-Parkin2025}. Our results further demonstrate that competitive $H_0$ measurements will be achievable with an O5-like configuration with moderately complete spectroscopic surveys, and that host weight mismodeling does not necessarily introduce significant systematics on $H_0$ even at moderate completeness levels.

These findings are particularly relevant for ongoing galaxy surveys such as \emph{Euclid} \citep{EuclidCollaboration2025}, DESI \citep{DESICollaboration2024}, and the proposed WST mission \citep{Mainieri2024}. With the increasing number of GW detections, it will be crucial to understand the interplay between the GW sources and host galaxy properties. Looking ahead, third-generation GW detectors such as the Einstein Telescope \citep{Punturo2010, Abac2025} and Cosmic Explorer \citep{Reitze2019} will be transformative in terms of detections, potentially revealing more detailed structures in the GW population and correlation with host galaxy properties \citep{Chen2024}. At the same time, space-based detectors such as LISA \citep{AmaroSeoane2023} will allow us to test these correlations at completely different black hole scales. Together, these advancements will be crucial to unlock the full potential of GWs as cosmological probes.

\begin{acknowledgements}
    We thank Michele Mancarella and Lucia Pozzetti for helpful suggestions and comments. 
    NB and MM acknowledge support from the grant ASI n. 2024-10-HH.0 ``Attività scientifiche per la missione \emph{Euclid} – fase E''. MM acknowledges support from MIUR, PRIN 2022 (grant 2022NY2ZRS 001). MM acknowledge the financial contribution from the grant PRIN-MUR 2022 2022NY2ZRS 001 "Optimizing the extraction of cosmological information from Large Scale Structure analysis in view of the next large spectroscopic surveys" supported by Next Generation EU. MT acknowledges the funding from the European Union - NextGenerationEU, in the framework of the HPC project – ``National Center for HPC, Big Data and Quantum Computing'' (PNRR - M4C2 - I1.4 - CN00000013 – CUP J33C22001170001). GC is supported by ERC Starting Grant No.~945155--GWmining, Cariplo Foundation Grant No.~2021-0555, MUR PRIN Grant No.~2022-Z9X4XS, Italian-French University (UIF/UFI) Grant No.~2025-C3-386, MUR Grant ``Progetto Dipartimenti di Eccellenza 2023-2027'' (BiCoQ), and the ICSC National Research Centre funded by NextGenerationEU. This material is based upon work supported by NSF's LIGO Laboratory which is a major facility fully funded by the National Science Foundation. We also acknowledge the CINECA award under the ISCRA initiative for the availability of high performance computing resources through ISCRA-C project LIGEA HP10CXW8OV and HP10CD9GBT (PI N. Borghi). \emph{Software:} \texttt{CHIMERA} \citep{Borghi2024, Tagliazucchi2025}, \texttt{GWFAST} \citep{Iacovelli2022_GWFAST}, \texttt{numpy} \citep{numpy}, \texttt{JAX} \citep{JAX}, \texttt{emcee} \citep{emcee}, \texttt{matplotlib} \citep{Hunter2007}, \texttt{datashader} \citep{datashader}, \texttt{pastamarkers} \citep{pastamarkers}.
\end{acknowledgements}

\bibliography{references}

\begin{thebibliography}{101}
\expandafter\ifx\csname natexlab\endcsname\relax\def\natexlab#1{#1}\fi

\bibitem[{Abac {et~al.}(2025{\natexlab{a}})Abac, Abramo, Albanesi, Albertini, Agapito, {et~al.}}]{Abac2025}
Abac, A., Abramo, R., Albanesi, S., {et~al.} 2025{\natexlab{a}}, [arXiv: \href{https://arxiv.org/abs/2503.12263}{\color{cobalt}2503.12263}]

\bibitem[{Abac {et~al.}(2025{\natexlab{b}})Abac, Abouelfettouh, Acernese, Ackley, Adamcewicz, {et~al.}}]{GWTC4_Cosmo}
Abac, A.~G., Abouelfettouh, I., Acernese, F., {et~al.} 2025{\natexlab{b}}, [arXiv: \href{https://arxiv.org/abs/2509.04348}{\color{cobalt}2509.04348}]

\bibitem[{Abac {et~al.}(2025{\natexlab{c}})Abac, Abouelfettouh, Acernese, Ackley, Adamcewicz, {et~al.}}]{GWTC4_Population}
Abac, A.~G., Abouelfettouh, I., Acernese, F., {et~al.} 2025{\natexlab{c}}, [arXiv: \href{https://arxiv.org/abs/2508.18083}{\color{cobalt}2508.18083}]

\bibitem[{Abac {et~al.}(2025{\natexlab{d}})Abac, Abouelfettouh, Acernese, Ackley, Adamcewicz, {et~al.}}]{GWTC4_Catalog}
Abac, A.~G., Abouelfettouh, I., Acernese, F., {et~al.} 2025{\natexlab{d}}, [arXiv: \href{https://arxiv.org/abs/2508.18082}{\color{cobalt}2508.18082}]

\bibitem[{Abbott {et~al.}(2016)Abbott, Abbott, Abbott, Abernathy, Acernese, {et~al.}}]{Abbott2016_Prospects}
Abbott, B.~P., Abbott, R., Abbott, T.~D., {et~al.} 2016, \href{http://dx.doi.org/10.1007/lrr-2016-1}{\color{cobalt}Living Reviews in Relativity}, 19, 1

\bibitem[{Abbott {et~al.}(2019)Abbott, Abbott, Abbott, Abraham, Acernese, {et~al.}}]{GWTC1_Catalog}
Abbott, B.~P., Abbott, R., Abbott, T.~D., {et~al.} 2019, \href{http://dx.doi.org/10.1103/physrevx.9.031040}{\color{cobalt}Physical Review X}, 9, 31040

\bibitem[{Abbott {et~al.}(2017{\natexlab{a}})Abbott, Abbott, Abbott, Acernese, Ackley, {et~al.}}]{Abbott2017_H0}
Abbott, B.~P., Abbott, R., Abbott, T.~D., {et~al.} 2017{\natexlab{a}}, \href{http://dx.doi.org/10.1038/nature24471}{\color{cobalt}\nat}, 551, 85

\bibitem[{Abbott {et~al.}(2017{\natexlab{b}})Abbott, Abbott, Abbott, Acernese, Ackley, {et~al.}}]{Abbott2017_GW170817}
Abbott, B.~P., Abbott, R., Abbott, T.~D., {et~al.} 2017{\natexlab{b}}, \href{http://dx.doi.org/10.1103/physrevlett.119.161101}{\color{cobalt}\prl}, 119, 161101

\bibitem[{Abbott {et~al.}(2021)Abbott, Abbott, Abraham, Acernese, Ackley, {et~al.}}]{GWTC2_Catalog}
Abbott, R., Abbott, T.~D., Abraham, S., {et~al.} 2021, \href{http://dx.doi.org/10.1103/physrevx.11.021053}{\color{cobalt}Physical Review X}, 11, 21053

\bibitem[{Abbott {et~al.}(2023{\natexlab{a}})Abbott, Abbott, Acernese, Ackley, Adams, {et~al.}}]{GWTC3_Catalog}
Abbott, R., Abbott, T.~D., Acernese, F., {et~al.} 2023{\natexlab{a}}, \href{http://dx.doi.org/10.1103/PhysRevX.13.041039}{\color{cobalt}Physical Review X}, 13, 41039

\bibitem[{Abbott {et~al.}(2023{\natexlab{b}})Abbott, Abbott, Acernese, Ackley, Adams, {et~al.}}]{Abbott2023_Population}
Abbott, R., Abbott, T.~D., Acernese, F., {et~al.} 2023{\natexlab{b}}, \href{http://dx.doi.org/10.1103/PhysRevX.13.011048}{\color{cobalt}Physical Review X}, 13, 011048

\bibitem[{Abbott {et~al.}(2023{\natexlab{c}})Abbott, Abe, Acernese, Ackley, Adhikari, {et~al.}}]{Abbott2023_Cosmo}
Abbott, R., Abe, H., Acernese, F., {et~al.} 2023{\natexlab{c}}, \href{http://dx.doi.org/10.3847/1538-4357/ac74bb}{\color{cobalt}\apj}, 949, 76

\bibitem[{Acernese {et~al.}(2015)Acernese, Agathos, Agatsuma, Aisa, Allemandou, {et~al.}}]{Acernese2015}
Acernese, F., Agathos, M., Agatsuma, K., {et~al.} 2015, \href{http://dx.doi.org/10.1088/0264-9381/32/2/024001}{\color{cobalt}Classical and Quantum Gravity}, 32, 24001

\bibitem[{Agarwal {et~al.}(2025)Agarwal, Dupletsa, Leyde, Mukherjee, Revenu, {et~al.}}]{Agarwal2025}
Agarwal, A., Dupletsa, U., Leyde, K., {et~al.} 2025, \href{http://dx.doi.org/10.3847/1538-4357/adda3a}{\color{cobalt}\apj}, 987, 47

\bibitem[{Alfradique {et~al.}(2025)Alfradique, Bom, \& Castro}]{Alfradique2025}
Alfradique, V., Bom, C.~R., \& Castro, T. 2025, \href{http://dx.doi.org/10.1103/vd36-3mys}{\color{cobalt}\prd}, 112, 063561

\bibitem[{Amaro-Seoane {et~al.}(2023)Amaro-Seoane, Andrews, Arca~Sedda, Askar, Baghi, {et~al.}}]{AmaroSeoane2023}
Amaro-Seoane, P., Andrews, J., Arca~Sedda, M., {et~al.} 2023, \href{http://dx.doi.org/10.1007/s41114-022-00041-y}{\color{cobalt}Living Reviews in Relativity}, 26, 2

\bibitem[{Artale {et~al.}(2020)Artale, Bouffanais, Mapelli, Giacobbo, Sabha, {et~al.}}]{Artale2020}
Artale, M.~C., Bouffanais, Y., Mapelli, M., {et~al.} 2020, \href{http://dx.doi.org/10.1093/mnras/staa1252}{\color{cobalt}\mnras}, 495, 1841

\bibitem[{Aso {et~al.}(2013)Aso, Michimura, Somiya, Ando, Miyakawa, {et~al.}}]{Aso2013}
Aso, Y., Michimura, Y., Somiya, K., {et~al.} 2013, \href{http://dx.doi.org/10.1103/PhysRevD.88.043007}{\color{cobalt}\prd}, 88, 43007

\bibitem[{Bednar {et~al.}(2025)Bednar, Crail, Thomas, {Jim Crist-Harif}, Rudiger, {et~al.}}]{datashader}
Bednar, J.~A., Crail, J., Thomas, I., {et~al.} 2025

\bibitem[{Borghi {et~al.}(2024)Borghi, Mancarella, Moresco, Tagliazucchi, Iacovelli, {et~al.}}]{Borghi2024}
Borghi, N., Mancarella, M., Moresco, M., {et~al.} 2024, \href{http://dx.doi.org/10.3847/1538-4357/ad20eb}{\color{cobalt}\apj}, 964, 191

\bibitem[{Bradbury {et~al.}(2018)Bradbury, Frostig, Hawkins, Johnson, Leary, {et~al.}}]{JAX}
Bradbury, J., Frostig, R., Hawkins, P., {et~al.} 2018, \url{http://github.com/jax-ml/jax}

\bibitem[{Bulla {et~al.}(2022)Bulla, Coughlin, Dhawan, \& Dietrich}]{Bulla2022}
Bulla, M., Coughlin, M.~W., Dhawan, S., \& Dietrich, T. 2022, \href{http://dx.doi.org/10.3390/universe8050289}{\color{cobalt}Universe}, 8, 289

\bibitem[{Callister(2024)}]{Callister2024_Review}
Callister, T.~A. 2024, [arXiv: \href{https://arxiv.org/abs/2410.19145}{\color{cobalt}2410.19145}]

\bibitem[{Carretero {et~al.}(2015)Carretero, Castander, Gazta{\~n}aga, Crocce, \& Fosalba}]{Carretero2015}
Carretero, J., Castander, F.~J., Gazta{\~n}aga, E., Crocce, M., \& Fosalba, P. 2015, \href{http://dx.doi.org/10.1093/mnras/stu2402}{\color{cobalt}\mnras}, 447, 646

\bibitem[{Chen {et~al.}(2024)Chen, Ezquiaga, \& Gupta}]{Chen2024}
Chen, H.-Y., Ezquiaga, J.~M., \& Gupta, I. 2024, \href{http://dx.doi.org/10.1088/1361-6382/ad424f}{\color{cobalt}Classical and Quantum Gravity}, 41, 125004

\bibitem[{Chen {et~al.}(2018)Chen, Fishbach, \& Holz}]{Chen2018}
Chen, H.-Y., Fishbach, M., \& Holz, D.~E. 2018, \href{http://dx.doi.org/10.1038/s41586-018-0606-0}{\color{cobalt}\nat}, 562, 545

\bibitem[{Chernoff \& Finn(1993)}]{Chernoff1993}
Chernoff, D.~F. \& Finn, L.~S. 1993, \href{http://dx.doi.org/10.1086/186898}{\color{cobalt}\apj}, 411, L5

\bibitem[{Collaboration {et~al.}(2024)Collaboration, Borghi, Ceccarelli, Croce, Leuzzi, {et~al.}}]{pastamarkers}
Collaboration, {\relax PASTA}., Borghi, N., Ceccarelli, E., {et~al.} 2024, [arXiv: \href{https://arxiv.org/abs/2403.20314}{\color{cobalt}2403.20314}]

\bibitem[{Crocce {et~al.}(2015)Crocce, Castander, Gazta{\~n}aga, Fosalba, \& Carretero}]{Crocce2015}
Crocce, M., Castander, F.~J., Gazta{\~n}aga, E., Fosalba, P., \& Carretero, J. 2015, \href{http://dx.doi.org/10.1093/mnras/stv1708}{\color{cobalt}\mnras}, 453, 1513

\bibitem[{{Cross-Parkin} {et~al.}(2025){Cross-Parkin}, Howlett, Davis, \& Khetan}]{Cross-Parkin2025}
{Cross-Parkin}, M.~L., Howlett, C., Davis, T.~M., \& Khetan, N. 2025, \href{http://dx.doi.org/10.1017/pasa.2025.10111}{\color{cobalt}\pasa}, 42, e149

\bibitem[{Dalang \& Baker(2024)}]{Dalang2024a}
Dalang, C. \& Baker, T. 2024, \href{http://dx.doi.org/10.1088/1475-7516/2024/02/024}{\color{cobalt}\jcap}, 2024, 24

\bibitem[{Del~Pozzo(2012)}]{DelPozzo2012}
Del~Pozzo, W. 2012, \href{http://dx.doi.org/10.1103/physrevd.86.043011}{\color{cobalt}\prd}, 86, 43011

\bibitem[{{DESI Collaboration} {et~al.}(2024){DESI Collaboration}, Adame, Aguilar, Ahlen, Alam, {et~al.}}]{DESICollaboration2024}
{DESI Collaboration}, Adame, A.~G., Aguilar, J., {et~al.} 2024, \href{http://dx.doi.org/10.3847/1538-3881/ad0b08}{\color{cobalt}\aj}, 167, 62

\bibitem[{Driver {et~al.}(2011)Driver, Hill, Kelvin, Robotham, Liske, {et~al.}}]{Driver2011}
Driver, S.~P., Hill, D.~T., Kelvin, L.~S., {et~al.} 2011, \href{http://dx.doi.org/10.1111/j.1365-2966.2010.18188.x}{\color{cobalt}\mnras}, 413, 971

\bibitem[{Edelman {et~al.}(2022)Edelman, Doctor, Godfrey, \& Farr}]{Edelman2022}
Edelman, B., Doctor, Z., Godfrey, J., \& Farr, B. 2022, \href{http://dx.doi.org/10.3847/1538-4357/ac3667}{\color{cobalt}\apj}, 924, 101

\bibitem[{{Euclid Collaboration} {et~al.}(2025){Euclid Collaboration}, Mellier, {Abdurro'uf}, Acevedo~Barroso, Achúcarro, {et~al.}}]{EuclidCollaboration2025}
{Euclid Collaboration}, Mellier, Y., {Abdurro'uf}, {et~al.} 2025, \href{http://dx.doi.org/10.1051/0004-6361/202450810}{\color{cobalt}\aap}, 697, A1

\bibitem[{Ezquiaga \& Holz(2022)}]{Ezquiaga2022}
Ezquiaga, J.~M. \& Holz, D.~E. 2022, \href{http://dx.doi.org/10.1103/physrevlett.129.061102}{\color{cobalt}\prl}, 129, 61102

\bibitem[{Farah {et~al.}(2025)Farah, Callister, Ezquiaga, Zevin, \& Holz}]{Farah2025}
Farah, A.~M., Callister, T.~A., Ezquiaga, J.~M., Zevin, M., \& Holz, D.~E. 2025, \href{http://dx.doi.org/10.3847/1538-4357/ad9253}{\color{cobalt}\apj}, 978, 153

\bibitem[{Farr {et~al.}(2019)Farr, Fishbach, Ye, \& Holz}]{Farr2019}
Farr, W.~M., Fishbach, M., Ye, J., \& Holz, D.~E. 2019, \href{http://dx.doi.org/10.3847/2041-8213/ab4284}{\color{cobalt}\apjl}, 883, L42

\bibitem[{Ferri {et~al.}(2025)Ferri, Tashiro, Abramo, Matos, Quartin, {et~al.}}]{Ferri2025}
Ferri, J., Tashiro, I.~L., Abramo, L.~R., {et~al.} 2025, \href{http://dx.doi.org/10.1088/1475-7516/2025/04/008}{\color{cobalt}\jcap}, 2025, 008

\bibitem[{Finke {et~al.}(2021)Finke, Foffa, Iacovelli, Maggiore, \& Mancarella}]{Finke2021_DarkSirens}
Finke, A., Foffa, S., Iacovelli, F., Maggiore, M., \& Mancarella, M. 2021, \href{http://dx.doi.org/10.1088/1475-7516/2021/08/026}{\color{cobalt}\jcap}, 2021, 26

\bibitem[{Fishbach {et~al.}(2019)Fishbach, Gray, Hernandez, Qi, Sur, {et~al.}}]{Fishbach2019_DarkGW170817}
Fishbach, M., Gray, R., Hernandez, I.~M., {et~al.} 2019, \href{http://dx.doi.org/10.3847/2041-8213/aaf96e}{\color{cobalt}\apjl}, 871, L13

\bibitem[{{Foreman-Mackey} {et~al.}(2013){Foreman-Mackey}, Hogg, Lang, \& Goodman}]{emcee}
{Foreman-Mackey}, D., Hogg, D.~W., Lang, D., \& Goodman, J. 2013, \href{http://dx.doi.org/10.1086/670067}{\color{cobalt}\pasp}, 125, 306

\bibitem[{Fosalba {et~al.}(2015{\natexlab{a}})Fosalba, Crocce, Gazta{\~n}aga, \& Castander}]{Fosalba2015a}
Fosalba, P., Crocce, M., Gazta{\~n}aga, E., \& Castander, F.~J. 2015{\natexlab{a}}, \href{http://dx.doi.org/10.1093/mnras/stv138}{\color{cobalt}\mnras}, 448, 2987

\bibitem[{Fosalba {et~al.}(2015{\natexlab{b}})Fosalba, Gazta{\~n}aga, Castander, \& Crocce}]{Fosalba2015b}
Fosalba, P., Gazta{\~n}aga, E., Castander, F.~J., \& Crocce, M. 2015{\natexlab{b}}, \href{http://dx.doi.org/10.1093/mnras/stu2464}{\color{cobalt}\mnras}, 447, 1319

\bibitem[{Gair {et~al.}(2023)Gair, Ghosh, Gray, Holz, Mastrogiovanni, {et~al.}}]{Gair2023}
Gair, J.~R., Ghosh, A., Gray, R., {et~al.} 2023, \href{http://dx.doi.org/10.3847/1538-3881/acca78}{\color{cobalt}\aj}, 166, 22

\bibitem[{Gorski {et~al.}(2005)Gorski, Hivon, Banday, Wandelt, Hansen, {et~al.}}]{Gorski2005}
Gorski, K.~M., Hivon, E., Banday, A.~J., {et~al.} 2005, \href{http://dx.doi.org/10.1086/427976}{\color{cobalt}\apj}, 622, 759

\bibitem[{Gray {et~al.}(2023)Gray, Beirnaert, Karathanasis, Revenu, Turski, {et~al.}}]{Gray2023}
Gray, R., Beirnaert, F., Karathanasis, C., {et~al.} 2023, \href{http://dx.doi.org/10.1088/1475-7516/2023/12/023}{\color{cobalt}\jcap}, 2023, 23

\bibitem[{Gray {et~al.}(2020)Gray, Hernandez, Qi, Sur, Brady, {et~al.}}]{Gray2020}
Gray, R., Hernandez, I.~M., Qi, H., {et~al.} 2020, \href{http://dx.doi.org/10.1103/physrevd.101.122001}{\color{cobalt}\prd}, 101, 122001

\bibitem[{Guzzo {et~al.}(2014)Guzzo, Scodeggio, Garilli, Granett, Fritz, {et~al.}}]{Guzzo2014}
Guzzo, L., Scodeggio, M., Garilli, B., {et~al.} 2014, \href{http://dx.doi.org/10.1051/0004-6361/201321489}{\color{cobalt}\aap}, 566, A108

\bibitem[{Hanselman {et~al.}(2025)Hanselman, Vijaykumar, Fishbach, \& Holz}]{Hanselman2025}
Hanselman, A.~G., Vijaykumar, A., Fishbach, M., \& Holz, D.~E. 2025, \href{http://dx.doi.org/10.3847/1538-4357/ad9393}{\color{cobalt}\apj}, 979, 9

\bibitem[{Harris {et~al.}(2020)Harris, Millman, Van Der~Walt, Gommers, Virtanen, {et~al.}}]{numpy}
Harris, C.~R., Millman, K.~J., Van Der~Walt, S.~J., {et~al.} 2020, \href{http://dx.doi.org/10.1038/s41586-020-2649-2}{\color{cobalt}\nat}, 585, 357

\bibitem[{Hoffmann {et~al.}(2015)Hoffmann, Bel, Gazta{\~n}aga, Crocce, Fosalba, {et~al.}}]{Hoffmann2015}
Hoffmann, K., Bel, J., Gazta{\~n}aga, E., {et~al.} 2015, \href{http://dx.doi.org/10.1093/mnras/stu2492}{\color{cobalt}\mnras}, 447, 1724

\bibitem[{Holz \& Hughes(2005)}]{Holz2005}
Holz, D.~E. \& Hughes, S.~A. 2005, \href{http://dx.doi.org/10.1086/431341}{\color{cobalt}\apj}, 629, 15

\bibitem[{Hotokezaka {et~al.}(2019)Hotokezaka, Nakar, Gottlieb, Nissanke, Masuda, {et~al.}}]{Hotokezaka2019}
Hotokezaka, K., Nakar, E., Gottlieb, O., {et~al.} 2019, \href{http://dx.doi.org/10.1038/s41550-019-0820-1}{\color{cobalt}Nature Astronomy}, 3, 940

\bibitem[{Hunter(2007)}]{Hunter2007}
Hunter, J.~D. 2007, \href{http://dx.doi.org/10.1109/MCSE.2007.55}{\color{cobalt}Computing in Science and Engineering}, 9, 90

\bibitem[{Iacovelli {et~al.}(2022{\natexlab{a}})Iacovelli, Mancarella, Foffa, \& Maggiore}]{Iacovelli2022}
Iacovelli, F., Mancarella, M., Foffa, S., \& Maggiore, M. 2022{\natexlab{a}}, \href{http://dx.doi.org/10.3847/1538-4357/ac9cd4}{\color{cobalt}\apj}, 941, 208

\bibitem[{Iacovelli {et~al.}(2022{\natexlab{b}})Iacovelli, Mancarella, Foffa, \& Maggiore}]{Iacovelli2022_GWFAST}
Iacovelli, F., Mancarella, M., Foffa, S., \& Maggiore, M. 2022{\natexlab{b}}, \href{http://dx.doi.org/10.3847/1538-4365/ac9129}{\color{cobalt}\apjs}, 263, 2

\bibitem[{Kalaghatgi {et~al.}(2020)Kalaghatgi, Hannam, \& Raymond}]{Kalaghatgi2020}
Kalaghatgi, C., Hannam, M., \& Raymond, V. 2020, \href{http://dx.doi.org/10.1103/physrevd.101.103004}{\color{cobalt}\prd}, 101, 103004

\bibitem[{Leja {et~al.}(2020)Leja, Speagle, Johnson, Conroy, van Dokkum, {et~al.}}]{Leja2020_SMF}
Leja, J., Speagle, J.~S., Johnson, B.~D., {et~al.} 2020, \href{http://dx.doi.org/10.3847/1538-4357/ab7e27}{\color{cobalt}\apj}, 893, 111

\bibitem[{Leyde {et~al.}(2024)Leyde, Baker, \& Enzi}]{Leyde2024}
Leyde, K., Baker, T., \& Enzi, W. 2024, \href{http://dx.doi.org/10.1088/1475-7516/2024/12/013}{\color{cobalt}\jcap}, 2024, 13

\bibitem[{Leyde {et~al.}(2025)Leyde, Baker, \& Enzi}]{Leyde2025}
Leyde, K., Baker, T., \& Enzi, W. 2025, [arXiv: \href{https://arxiv.org/abs/2507.12171}{\color{cobalt}2507.12171}]

\bibitem[{Leyde {et~al.}(2022)Leyde, Mastrogiovanni, Steer, Chassande-Mottin, \& Karathanasis}]{Leyde2022}
Leyde, K., Mastrogiovanni, S., Steer, D., Chassande-Mottin, E., \& Karathanasis, C. 2022, \href{http://dx.doi.org/10.1088/1475-7516/2022/09/012}{\color{cobalt}\jcap}, 2022, 12

\bibitem[{{LIGO Scientific Collaboration} {et~al.}(2015){LIGO Scientific Collaboration}, Aasi, Abbott, Abbott, Abbott, {et~al.}}]{LIGOScientificCollaboration2015}
{LIGO Scientific Collaboration}, Aasi, J., Abbott, B.~P., {et~al.} 2015, \href{http://dx.doi.org/10.1088/0264-9381/32/7/074001}{\color{cobalt}Classical and Quantum Gravity}, 32, 74001

\bibitem[{Lilly {et~al.}(2007)Lilly, Le~Fèvre, Renzini, Zamorani, Scodeggio, {et~al.}}]{Lilly2007}
Lilly, S.~J., Le~Fèvre, O., Renzini, A., {et~al.} 2007, \href{http://dx.doi.org/10.1086/516589}{\color{cobalt}\apjs}, 172, 70

\bibitem[{London {et~al.}(2018)London, Khan, {Fauchon-Jones}, Garc{\'i}a, Hannam, {et~al.}}]{London2018}
London, L., Khan, S., {Fauchon-Jones}, E., {et~al.} 2018, \href{http://dx.doi.org/10.1103/physrevlett.120.161102}{\color{cobalt}\prl}, 120, 161102

\bibitem[{Loredo(2004)}]{Loredo2004}
Loredo, T.~J. 2004, \href{http://dx.doi.org/10.1063/1.1835214}{\color{cobalt}AIP Conference Proceedings}, 735, 195

\bibitem[{Madau \& Dickinson(2014)}]{Madau2014}
Madau, P. \& Dickinson, M. 2014, \href{http://dx.doi.org/10.1146/annurev-astro-081811-125615}{\color{cobalt}\araa}, 52, 415

\bibitem[{Mainieri {et~al.}(2024)Mainieri, Anderson, Brinchmann, Cimatti, Ellis, {et~al.}}]{Mainieri2024}
Mainieri, V., Anderson, R.~I., Brinchmann, J., {et~al.} 2024, [arXiv: \href{https://arxiv.org/abs/2403.05398}{\color{cobalt}2403.05398}]

\bibitem[{Mali \& Essick(2025)}]{Mali2025}
Mali, U. \& Essick, R. 2025, \href{http://dx.doi.org/10.3847/1538-4357/ad9de7}{\color{cobalt}\apj}, 980, 85

\bibitem[{Mancarella {et~al.}(2022)Mancarella, {Genoud-Prachex}, \& Maggiore}]{Mancarella2022}
Mancarella, M., {Genoud-Prachex}, E., \& Maggiore, M. 2022, \href{http://dx.doi.org/10.1103/physrevd.105.064030}{\color{cobalt}\prd}, 105, 64030

\bibitem[{Mandel {et~al.}(2019)Mandel, Farr, \& Gair}]{Mandel2019}
Mandel, I., Farr, W.~M., \& Gair, J.~R. 2019, \href{http://dx.doi.org/10.1093/mnras/stz896}{\color{cobalt}\mnras}, 486, 1086

\bibitem[{Mastrogiovanni {et~al.}(2023)Mastrogiovanni, Laghi, Gray, Santoro, Ghosh, {et~al.}}]{Mastrogiovanni2023}
Mastrogiovanni, S., Laghi, D., Gray, R., {et~al.} 2023, \href{http://dx.doi.org/10.1103/physrevd.108.042002}{\color{cobalt}\prd}, 108, 42002

\bibitem[{Mastrogiovanni {et~al.}(2021)Mastrogiovanni, Leyde, Karathanasis, Chassande-Mottin, Steer, {et~al.}}]{Mastrogiovanni2021_icarogw}
Mastrogiovanni, S., Leyde, K., Karathanasis, C., {et~al.} 2021, \href{http://dx.doi.org/10.1103/PhysRevD.104.062009}{\color{cobalt}\prd}, 104, 062009

\bibitem[{Mastrogiovanni {et~al.}(2024)Mastrogiovanni, Pierra, Perriès, Laghi, Santoro, {et~al.}}]{Mastrogiovanni2024_icarogw}
Mastrogiovanni, S., Pierra, G., Perriès, S., {et~al.} 2024, \href{http://dx.doi.org/10.1051/0004-6361/202347007}{\color{cobalt}\aap}, 682, A167

\bibitem[{Moresco {et~al.}(2022)Moresco, Amati, Amendola, Birrer, Blakeslee, {et~al.}}]{Moresco2022}
Moresco, M., Amati, L., Amendola, L., {et~al.} 2022, \href{http://dx.doi.org/10.1007/s41114-022-00040-z}{\color{cobalt}Living Reviews in Relativity}, 25, 6

\bibitem[{Mukherjee {et~al.}(2021)Mukherjee, Wandelt, Nissanke, \& Silvestri}]{Mukherjee2021}
Mukherjee, S., Wandelt, B.~D., Nissanke, S.~M., \& Silvestri, A. 2021, \href{http://dx.doi.org/10.1103/PhysRevD.103.043520}{\color{cobalt}\prd}, 103, 043520

\bibitem[{Oguri(2016)}]{Oguri2016}
Oguri, M. 2016, \href{http://dx.doi.org/10.1103/physrevd.93.083511}{\color{cobalt}\prd}, 93, 83511

\bibitem[{Palmese {et~al.}(2020)Palmese, deVicente, Pereira, Annis, Hartley, {et~al.}}]{Palmese2020}
Palmese, A., deVicente, J., Pereira, M. E.~S., {et~al.} 2020, \href{http://dx.doi.org/10.3847/2041-8213/abaeff}{\color{cobalt}\apjl}, 900, L33

\bibitem[{Palmese {et~al.}(2024)Palmese, Kaur, Hajela, Margutti, McDowell, {et~al.}}]{Palmese2024}
Palmese, A., Kaur, R., Hajela, A., {et~al.} 2024, \href{http://dx.doi.org/10.1103/PhysRevD.109.063508}{\color{cobalt}\prd}, 109, 063508

\bibitem[{Palmese \& Mastrogiovanni(2025)}]{Palmese2025}
Palmese, A. \& Mastrogiovanni, S. 2025, [arXiv: \href{https://arxiv.org/abs/2502.00239}{\color{cobalt}2502.00239}]

\bibitem[{Pedrotti {et~al.}(2025)Pedrotti, Mancarella, Bel, \& Gerosa}]{Pedrotti2025}
Pedrotti, A., Mancarella, M., Bel, J., \& Gerosa, D. 2025, [arXiv: \href{https://arxiv.org/abs/2504.10482}{\color{cobalt}2504.10482}]

\bibitem[{Perna {et~al.}(2025)Perna, Mastrogiovanni, \& Ricciardone}]{Perna2025}
Perna, G., Mastrogiovanni, S., \& Ricciardone, A. 2025, \href{http://dx.doi.org/10.1051/0004-6361/202450840}{\color{cobalt}\aap}, 698, A128

\bibitem[{Pierra \& Mastrogiovanni(2025)}]{Pierra2025}
Pierra, G. \& Mastrogiovanni, S. 2025, [arXiv: \href{https://arxiv.org/abs/2507.10597}{\color{cobalt}2507.10597}]

\bibitem[{Pozzetti {et~al.}(2010)Pozzetti, Bolzonella, Zucca, Zamorani, Lilly, {et~al.}}]{Pozzetti2010}
Pozzetti, L., Bolzonella, M., Zucca, E., {et~al.} 2010, \href{http://dx.doi.org/10.1051/0004-6361/200913020}{\color{cobalt}\aap}, 523, A13

\bibitem[{Punturo {et~al.}(2010)Punturo, Abernathy, Acernese, Allen, Andersson, {et~al.}}]{Punturo2010}
Punturo, M., Abernathy, M., Acernese, F., {et~al.} 2010, \href{http://dx.doi.org/10.1088/0264-9381/27/19/194002}{\color{cobalt}Classical and Quantum Gravity}, 27, 194002

\bibitem[{Reitze {et~al.}(2019)Reitze, Adhikari, Ballmer, Barish, Barsotti, {et~al.}}]{Reitze2019}
Reitze, D., Adhikari, R.~X., Ballmer, S., {et~al.} 2019, \href{http://dx.doi.org/10.48550/ARXIV.1907.04833}{\color{cobalt}\baas}, 51, 35

\bibitem[{Santoliquido {et~al.}(2022)Santoliquido, Mapelli, Artale, \& Boco}]{Santoliquido2022}
Santoliquido, F., Mapelli, M., Artale, M.~C., \& Boco, L. 2022, \href{http://dx.doi.org/10.1093/mnras/stac2384}{\color{cobalt}\mnras}, 516, 3297

\bibitem[{Schechter(1976)}]{Schechter1976}
Schechter, P. 1976, \href{http://dx.doi.org/10.1086/154079}{\color{cobalt}\apj}, 203, 297

\bibitem[{Schutz(1986)}]{Schutz1986}
Schutz, B.~F. 1986, \href{http://dx.doi.org/10.1038/323310a0}{\color{cobalt}\nat}, 323, 310

\bibitem[{Tagliazucchi {et~al.}(2025)Tagliazucchi, Moresco, Borghi, \& Fiebig}]{Tagliazucchi2025}
Tagliazucchi, M., Moresco, M., Borghi, N., \& Fiebig, M. 2025, \href{http://dx.doi.org/10.1051/0004-6361/202554827}{\color{cobalt}\aap}, 702, A244

\bibitem[{Talbot \& Thrane(2018)}]{Talbot2018}
Talbot, C. \& Thrane, E. 2018, \href{http://dx.doi.org/10.3847/1538-4357/aab34c}{\color{cobalt}\apj}, 856, 173

\bibitem[{Taylor {et~al.}(2012)Taylor, Gair, \& Mandel}]{Taylor2012}
Taylor, S.~R., Gair, J.~R., \& Mandel, I. 2012, \href{http://dx.doi.org/10.1103/PhysRevD.85.023535}{\color{cobalt}\prd}, 85, 023535

\bibitem[{Turisini {et~al.}(2024)Turisini, Cestari, \& Amati}]{Turisini2024}
Turisini, M., Cestari, M., \& Amati, G. 2024, \href{http://dx.doi.org/10.17815/jlsrf-8-186}{\color{cobalt}Journal of Large-scale Research Facilities JLSRF}, 9

\bibitem[{Turski {et~al.}(2025)Turski, Brozzetti, D{\'a}lya, Punturo, \& Ghosh}]{Turski2025}
Turski, C., Brozzetti, M.~L., D{\'a}lya, G., Punturo, M., \& Ghosh, A. 2025, [arXiv: \href{https://arxiv.org/abs/2505.13568}{\color{cobalt}2505.13568}]

\bibitem[{Valentino {et~al.}(2025)Valentino, Said, Riess, Pollo, Poulin, {et~al.}}]{DiValentino2025}
Valentino, E.~D., Said, J.~L., Riess, A., {et~al.} 2025, \href{http://dx.doi.org/10.1016/j.dark.2025.101965}{\color{cobalt}Physics of the Dark Universe}, 49, 101965

\bibitem[{Verde {et~al.}(2019)Verde, Treu, \& Riess}]{Verde2019}
Verde, L., Treu, T., \& Riess, A.~G. 2019, \href{http://dx.doi.org/10.1038/s41550-019-0902-0}{\color{cobalt}Nature Astronomy}, 3, 891

\bibitem[{Vijaykumar {et~al.}(2024)Vijaykumar, Fishbach, Adhikari, \& Holz}]{Vijaykumar2024}
Vijaykumar, A., Fishbach, M., Adhikari, S., \& Holz, D.~E. 2024, \href{http://dx.doi.org/10.3847/1538-4357/ad6140}{\color{cobalt}\apj}, 972, 157

\bibitem[{Vitale {et~al.}(2022)Vitale, Gerosa, Farr, \& Taylor}]{Vitale2022}
Vitale, S., Gerosa, D., Farr, W.~M., \& Taylor, S.~R. 2022, [arXiv: \href{https://arxiv.org/abs/2007.05579}{\color{cobalt}2007.05579}]

\bibitem[{Weaver {et~al.}(2023)Weaver, Davidzon, Toft, Ilbert, McCracken, {et~al.}}]{Weaver2023}
Weaver, J.~R., Davidzon, I., Toft, S., {et~al.} 2023, \href{http://dx.doi.org/10.1051/0004-6361/202245581}{\color{cobalt}\aap}, 677, A184

\bibitem[{York {et~al.}(2000)York, Adelman, Anderson, Anderson, Annis, {et~al.}}]{York2000}
York, D.~G., Adelman, J., Anderson, Jr., J.~E., {et~al.} 2000, \href{http://dx.doi.org/10.1086/301513}{\color{cobalt}\aj}, 120, 1579

\end{thebibliography}

\begin{appendix}

\section{Mock catalogs generation}\label{app:mice_sch}

\begin{figure}
    \centering
    \includegraphics[width=1\hsize]{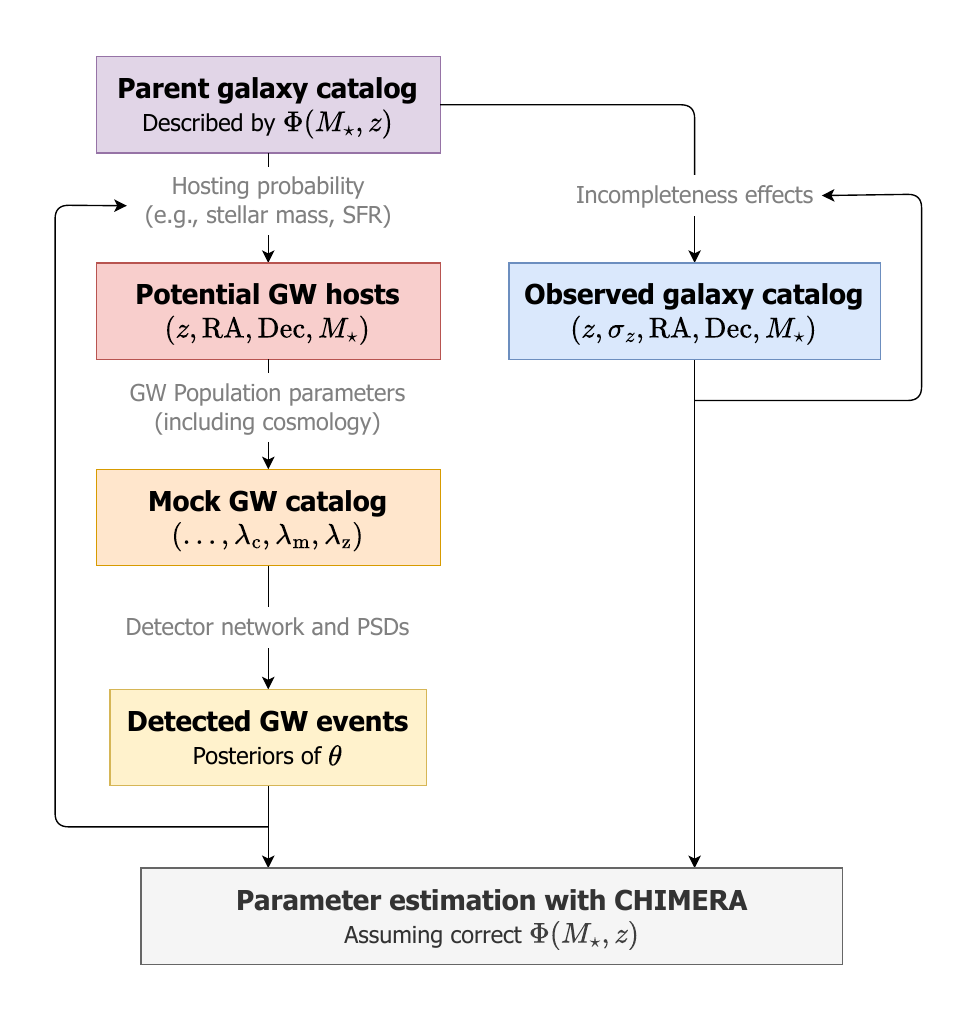}
    \caption{Overview of our end-to-end simulation framework for testing the impact of different galaxy catalog completeness cuts and host weighting schemes on standard siren constraints.}
    \label{fig:framework}
\end{figure}

Figure~\ref{fig:framework} summarizes the mock data generation framework presented in this work. The parent galaxy catalog used in this work is obtained from MICECATv2~\citep{Fosalba2015a,Fosalba2015b,Crocce2015,Carretero2015,Hoffmann2015}. To ensure an efficient computation of the galaxy term (Eq.~\ref{eq:pgal_final}) while preserving realistic density and clustering properties, the full catalog is subsampled and modeled assuming a theoretical stellar mass function (SMF). From this catalog, we generate three sub-catalogs of potential host galaxies varying the hosting probability exponent $\alpha_M$ as follows: unweighted ($\alpha_M = 0$), mass weighted ($\alpha_M = 1$), squared mass weighted ($\alpha_M = 2$). For each of these catalogs, we generate a GW population based on the prescriptions of Sect.~\ref{subsec:mockGW} and compute the detected events with a Fisher Matrix approach as described in Sect.~\ref{subsec:detGW}. These GW catalogs are analyzed in \texttt{CHIMERA} together with the galaxy catalog, taking into account catalog incompleteness cuts.

To describe the parent galaxy catalog, we adopt a double Schechter function \citep[e.g.,][]{Pozzetti2010, Leja2020_SMF}:
\begin{equation}\label{eq:schechter}
    \Phi(M,z) \, \mathrm{d}M = \begin{cases} 
                                \left[ \Phi_1^\ast(z) \left( \frac{M}{M^\ast} \right)^{\alpha_1}\!\! + \Phi_2^\ast(z) \left( \frac{M}{M^\ast} \right)^{\alpha_2} \right] \mathrm{e}^{-\frac{M}{M^\ast}} \frac{\mathrm{d}M}{M^\ast} & \text{if } M \geq M_{\rm lim}(z)\,,\\
                                0 & \text{otherwise}\,,
                                \end{cases}
\end{equation}
where $\Phi_{1,2}^\ast$ are the characteristic number densities of the two components, $\alpha_{1,2}$ are the low-mass slope parameters, and $M^\ast$ is the characteristic mass scale separating the power-law regime ($M \ll M^\ast$) from the exponential cutoff regime ($M \gg M^\ast$). $M_{\rm lim}$ is the stellar mass completeness limit, representing the minimum stellar mass above which the galaxy catalog is complete. We incorporate redshift evolution by allowing $\Phi_1^\ast$, $\Phi_2^\ast$, and $\log_{10} M_{\rm lim}$ to vary according to a quadratic function:
\begin{equation}\label{eq:schechter_params}
    p(z) = a_0 + a_1(1+z) + a_2(1+z)^2\,,
\end{equation}
while keeping $\log_{10} M_\ast$, $\alpha_1$, and $\alpha_2$ fixed. This parameterization ultimately provides a model consistent with recent observational constraints \citep[see][]{Leja2020_SMF,Weaver2023}. The parent galaxy catalog is then built as follows.
\begin{enumerate}
    \item We bin the original MICECATv2 catalog into 40 redshift bins spanning $0.07 < z < 1.4$ and 40 logarithmic stellar mass bins ranging from $6.5 < \log_{10}(M_\star/M_\odot) < 12.2$. Within each redshift bin, we compute the catalog SMF and define $M_{\rm lim}$ as the stellar mass corresponding to the peak number density (see Fig.~\ref{fig:mice_Sch}, left panel).
    
    \item We fit a quadratic function (Eq.~\ref{eq:schechter_params}) to $\log_{10} M_{\rm lim}(z)$ across redshift bins and remove all galaxies with a stellar mass below $M_{\rm lim}$ in each redshift bin. This is a first step to ensure that the resulting catalog is actually mass-complete above $M_{\rm lim}(z)$ and can be accurately described by the theoretical SMF.

    \item In each redshift bin, the catalog SMF is smoothed with a $1\,\sigma$ Gaussian filter to reduce the statistical fluctuations (Fig.~\ref{fig:mice_Sch}, central panel) and fitted with a double Schechter function following Eq.~(\ref{eq:schechter}), with redshift evolution modeled via the quadratic parameterization in Eq.~(\ref{eq:schechter_params}). To improve the fit convergence, we fix $M^\ast$ and slopes $\alpha_{1,2}$ to values consistent with the continuity model from \citet{Leja2020_SMF} applied to the COSMOS2020 catalog \citep{Weaver2023}. The resulting coefficients are provided in Table~\ref{tab:mice_Sch_coeffs}.
    
    \item At this point, we have a smooth parametric model of the MICECATv2 SMF across its full mass and redshift range. We validate it by comparing the catalog number density with the theoretical number density obtained by integrating $\Phi(M,z)$ (Fig.~\ref{fig:mice_Sch}, central panel). The average difference remains below $5\%$, with oscillations likely due to synthetic effects inherent in the simulation. This very good agreement demonstrates that our parametric SMF accurately represents the underlying galaxy population.
\end{enumerate}
The resulting catalog, referred to as the parent catalog throughout this paper, contains approximately 335 million galaxies, representing about 67\% of the full MICECATv2 catalog. This subsample is more than 200 times larger than the high-mass subsample of MICECATv2 galaxies used in \citet{Borghi2024}. 

The framework presented here has been implemented in \texttt{CHIMERA} (v2.1) with a vectorized approach to quickly evaluate the Schechter function (Eqs.~\ref{eq:schechter} and \ref{eq:schechter_params}) and its integral across multiple sky pixels. The catalog term and completeness functions can be pre-computed and stored for different sky masks (see Sect.~\ref{sec:method}). This means that with the above parameterization, it is possible to efficiently compute the background term (Eq.~\ref{eq:pbkg}) and thus the final galaxy term (Eq.~\ref{eq:pgal_final}).

\begin{table}
    \caption{Coefficients for the redshift-evolving double Schechter function describing the parent galaxy catalog. \label{tab:mice_Sch_coeffs}}
    \centering
    \begin{tabular}{cccc}
    \hline\hline
    Parameter & $a_0$ & $a_1$ & $a_2$ \\
    \hline
    $\log_{10} M_{\rm lim}$ & $2.90$ & $5.32$ & $-1.04$ \\
    $\log_{10} \Phi_1$ & $1.48 \times 10^{-3}$ & $3.44 \times 10^{-3}$ & $-1.68 \times 10^{-3}$ \\
    $\log_{10} \Phi_2$ & $3.23 \times 10^{-3}$ & $-2.40 \times 10^{-3}$ & $0.57 \times 10^{-3}$ \\
    $\log_{10} M^\ast$ & $10.5$ & $0$ &  $0$ \\
    $\alpha_1$ & $-0.3$ &  $0$ &  $0$ \\
    $\alpha_2$ & $-1.5$ &  $0$ &  $0$ \\
    \hline
    \end{tabular}
    \tablefoot{Schechter parameters $\{\log_{10} M^\ast, \alpha_1, \alpha_2\}$ are kept fixed in the fit.}
\end{table}

\begin{table*}
    \centering
    \caption{Constraints on $H_0$ from 100 BBHs in an LVK O5-like scenario for different galaxy catalog completeness levels and host weightings. \label{tab:res_H0}}
    \small
    \renewcommand{\arraystretch}{1.2}
    \begin{tabular*}{\textwidth}{@{\extracolsep{\fill}}l@{\hspace{8pt}}c@{\hspace{8pt}}c@{\hspace{8pt}}c@{\hspace{8pt}}c@{\hspace{8pt}}c@{\hspace{8pt}}c@{\hspace{8pt}}c@{\hspace{8pt}}c@{\hspace{8pt}}c@{}}
    \hline\hline
    \multirow{3}{*}{\textbf{Completeness}} & \multicolumn{3}{c}{\texttt{mock 0 -- $w \propto \mathrm{const.}$}} & \multicolumn{3}{c}{\texttt{mock 1 -- $w \propto \M$}} & \multicolumn{3}{c}{\texttt{mock 2 -- $w \propto \M^2$}} \\[0.2em]
    \cline{2-4} \cline{5-7} \cline{8-10}
     & $\langle P_{\rm compl}\rangle$ & $N_{\rm host}$ & $H_0$ & $\langle P_{\rm compl}\rangle$ & $N_{\rm host}$ & $H_0$ & $\langle P_{\rm compl}\rangle$ & $N_{\rm host}$ & $H_0$ \\
     & (\%) & & (km\,s$^{-1}$\,Mpc$^{-1}$) & (\%) & & (km\,s$^{-1}$\,Mpc$^{-1}$) & (\%) & & (km\,s$^{-1}$\,Mpc$^{-1}$) \\
    \hline
    Complete & 89.8 & 100 & $70.6^{+1.1}_{-1.1}~(1.6\%)$ & 89.6 & 100 & $70.3^{+0.9}_{-0.9}~(1.3\%)$ & 93.8 & 100 & $69.9^{+0.6}_{-0.6}~(0.9\%)$ \\[0.7em]
    $\log_{10}\,(M_\star/M_\odot) > 9.5$ & 51.5 & 26 & $70.9^{+1.5}_{-1.4}~(2.0\%)$ & 85.8 & 90 & $70.4^{+0.9}_{-0.9}~(1.3\%)$ & 93.8 & 99 & $70.1^{+0.6}_{-0.6}~(0.9\%)$ \\
    $\log_{10}\,(M_\star/M_\odot) > 10.0$ & 23.6 & 14 & $71.2^{+1.9}_{-1.8}~(2.6\%)$ & 71.9 & 76 & $70.6^{+0.9}_{-0.8}~(1.2\%)$ & 93.5 & 97 & $70.2^{+0.7}_{-0.6}~(0.9\%)$ \\
    $\log_{10}\,(M_\star/M_\odot) > 10.5$ & 6.94 & 5 & $70.9^{+3.4}_{-3.3}~(4.7\%)$ & 44.5 & 45 & $70.2^{+0.7}_{-0.7}~(1.0\%)$ & 89.5 & 86 & $70.2^{+0.6}_{-0.6}~(0.9\%)$ \\
    $\log_{10}\,(M_\star/M_\odot) > 11.0$  & 0.83 & 1 & $71.2^{+10.5}_{-6.7}~(12\%)$ & 12.6 & 13 & $70.7^{+2.6}_{-1.7}~(3.1\%)$ & 52.3 & 31 & $70.0^{+0.7}_{-0.7}~(1.0\%)$ \\
    $\log_{10}\,(M_\star/M_\odot) > 11.5$ & 0.01 & 0 & $71.2^{+13.9}_{-8.0}~(15\%)$ & 0.53 & 2 & $72.4^{+12.8}_{-5.7}~(13\%)$ & 5.56 & 1 & $68.9^{+1.8}_{-1.8}~(2.6\%)$ \\[0.7em]
    $i < 24$ & 65.6 & 84 & $70.5^{+1.1}_{-1.2}~(1.7\%)$ & 76.8 & 99 & $70.4^{+0.7}_{-0.7}~(1.1\%)$ & 85.2 & 100 & $70.2^{+0.6}_{-0.6}~(0.9\%)$ \\
    $i < 22$ & 14.4 & 34 & $70.5^{+1.5}_{-1.6}~(2.2\%)$ & 42.6 & 83 & $70.3^{+0.9}_{-0.8}~(1.2\%)$ & 53.5 & 89 & $70.0^{+0.6}_{-0.6}~(0.9\%)$ \\
    $i < 20$ & 2.63 & 10 & $69.9^{+3.0}_{-3.4}~(4.6\%)$ & 20.9 & 43 & $70.4^{+0.8}_{-0.8}~(1.1\%)$ & 30.8 & 68 & $70.0^{+0.7}_{-0.6}~(0.9\%)$  \\
    $i < 18$ & 0.40 & 0 & $69.9^{+9.2}_{-6.9}~(12\%)$ & 7.22 & 17 & $69.7^{+0.9}_{-0.8}~(1.2\%)$ & 15.2 & 33 & $70.0^{+0.6}_{-0.6}~(0.9\%)$ \\
    $i < 16$ & 0.03 & 0 & $70.9^{+13.4}_{-7.9}~(15\%)$ & 1.14 & 3 & $71.4^{+7.7}_{-1.3}~(6.3\%)$ & 4.17 & 2 & $77.0^{+6.9}_{-8.7}~(10\%)$ \\[0.7em]
    Empty/Spectral & 0 & 0 & $71.5^{+14.7}_{-8.5}~(16\%)$ & 0 & 0 & $77.0^{+13.8}_{-7.8}~(14\%)$ & 0 & 0 & $72.2^{+14.9}_{-10.9}~(18\%)$ \\
    \hline
    \end{tabular*}
    \tablefoot{$\langle P_{\rm compl} \rangle$ is a comoving volume-weighted average within the GW detector horizon, $0 < z < 1.4$.}
\end{table*}

\section{Results table}\label{app:res_table}
Table~\ref{tab:res_H0} reports the marginalized $H_0$ constraints at different completeness configurations for the three host galaxy weighting schemes studied in this work.
For each configuration, we provide the average completeness, $\langle P_{\rm compl}\rangle$, computed following Eq.~(\ref{eq:avg_Pcompl}). We stress that $P_{\rm compl}$ refers to potential BBH host galaxies not to the full galaxy population (see its definition in Eq.~\ref{eq:completeness_function}). We also include the number of actual host galaxies $N_{\rm host}$ remaining in the catalog after applying observational selection cuts.

\section{Impact of wrong host galaxy weighting}\label{app:syst}

\begin{figure}
    \centering
    \includegraphics[width=\linewidth]{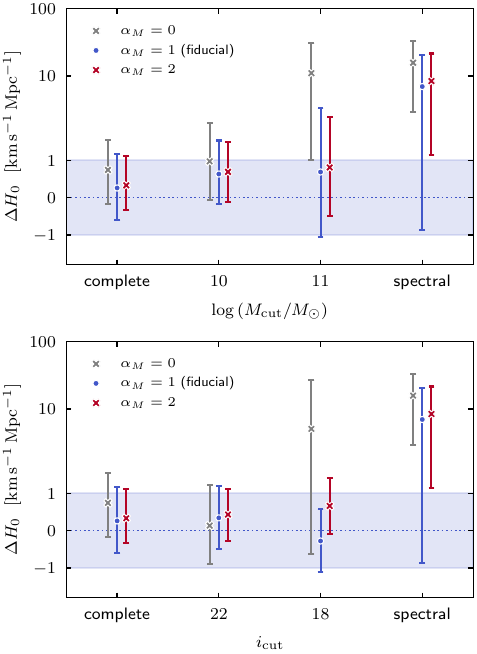}
    \caption{Bias on marginalized $H_0$ constraints at varying completeness levels for 100 BBH detections in an LVK O5-like scenario generated from a mass-weighted catalog (\texttt{mock 1}) assuming three host weighting schemes: constant (wrong), linear (fiducial), and squared (wrong) stellar mass weighting.}\label{fig:H0_bias}
\end{figure}

Figure~\ref{fig:H0_bias} shows the bias in marginalized $H_0$ constraints when incorrect host weighting assumptions are applied during parameter estimation. We test the mass-weighted \texttt{mock1} as our fiducial case across various completeness cuts. When correctly setting $\alpha_M = 1$, all configurations yield unbiased $H_0$ constraints within $1\,\sigma$. The shift in the median $H_0$ in the spectral siren regime is likely due to the specific realization of GW events. Therefore, wrong configurations are also expected to be skewed toward this value.

For complete spectroscopic catalogs, incorrect weighting introduces systematic errors that are negligible compared to statistical uncertainties (with $|\Delta H_0| < 1~\mathrm{km\,s^{-1}\,Mpc^{-1}}$). Interestingly, the mass-squared weighting ($\alpha_M = 2$) yields slightly tighter constraints compared to the true weighting. This indicates that for the configurations studied in this work, a stronger down-weighting of low-mass galaxies reduces the effective number of galaxies contributing to the catalog term without introducing substantial bias. 
As average completeness decreases below ${\approx}20\%$, using a stronger-than-true mass weighting still yields unbiased results. This is also expected as $\alpha_M = 1$ and $\alpha_M = 2$ configurations have similar background distributions (see Fig.~\ref{fig:mock_properties}, lower left panel). In contrast, using $\alpha_M = 0$ produces biases of about $10~\mathrm{km\,s^{-1}\,Mpc^{-1}}$. When catalogs become highly incomplete or empty (spectral siren regime), both incorrect weighting assumptions produce biases, exceeding $1\,\sigma$ for $\alpha_M = 0$ and $2\,\sigma$ for $\alpha_M = 2$.
 
\end{appendix}

\end{document}